\documentclass[fleqn,usenatbib]{mnras}
\usepackage{newtxtext,newtxmath}

\usepackage[T1]{fontenc}

\DeclareRobustCommand{\VAN}[3]{#2}
\let\VANthebibliography\thebibliography
\def\thebibliography{\DeclareRobustCommand{\VAN}[3]{##3}\VANthebibliography}

\usepackage{graphicx}	

\title[Constraints on NS ellipticity from SGWB]{Stochastic gravitational-wave background searches and constraints on neutron-star ellipticity}

\author[]{
Federico De Lillo,\thanks{E-mail: federico.delillo@uclouvain.be}
Jishnu Suresh,\thanks{E-mail: jishnu.suresh@uclouvain.be}
Andrew L. Miller\thanks{E-mail: andrew.miller@uclouvain.be}
\\
Centre for Cosmology, Particle Physics and Phenomenology (CP3),\\
Universit\'e catholique de Louvain, Louvain-la-Neuve, B-1348, Belgium
}

\date{Accepted XXX. Received YYY; in original form ZZZ}

\pubyear{2022}

\begin{document}
\label{firstpage}
\pagerange{\pageref{firstpage}--\pageref{lastpage}}
\maketitle

\begin{abstract}
Rotating neutron stars (NSs) are promising sources of gravitational waves (GWs) in the frequency band of ground-based detectors. They are expected to emit quasi-monochromatic, long-duration GW signals, called continuous waves (CWs), due to their deviations from spherical symmetry. The degree of such deformations, and hence the information about the internal structure of a NS, is encoded in a dimensionless parameter $\varepsilon$ called ellipticity. Searches for CW signals from isolated Galactic NSs have shown to be sensitive to ellipticities as low as $\varepsilon \sim \mathcal{O}(10^{-9})$. These searches are optimal for detecting and characterising GWs from individual NSs, but they are not designed to measure the properties of NSs as population, such as the average ellipticity $\varepsilon_{\mathrm{av}}$. These ensemble properties can be determined by the measurement of the stochastic gravitational-wave background (SGWB) arising from the superposition of GW signals from individually-undetectable NSs. In this work, we perform a cross-correlation search for such a SGWB using the data from the first three observation runs of Advanced LIGO and Virgo. Finding no evidence for a SGWB signal, we set upper limits on the dimensionless energy density parameter $\Omega_{\mathrm{gw}}(f)$. Using these results, we also constrain the average ellipticity of Galactic NSs and five NS ``hotspots'', as a function of the number of NSs emitting GWs within the frequency band of the search $N_{\mathrm{band}}$. We find $\varepsilon_{\mathrm{av}} \la 1.8 \times 10^{-8}$, with $N_{\mathrm{band}}=1.6 \times 10^7$, for Galactic NSs, and $\varepsilon_{\mathrm{av}} \la [3.5-11.8]\times 10^{-7}$, with $N_{\mathrm{band}}=1.6 \times 10^{10}$, for NS hotspots.
\end{abstract}



\section{Introduction}
\label{sec:introduction}

Isolated, rotating, non axi-symmetric neutron stars, with a rotational period of the order of milliseconds, are promising sources of GWs for ground-based GW detectors, such Advanced LIGO~\citep{2015CQGra..32g4001L}, Advanced Virgo~\citep{2015CQGra..32b4001A}, and KAGRA \citep{KAGRA:2020tym}. Such objects would emit GWs due to deformations on their surfaces, i.e. ``mountains'' \citep{Jones:2001yg}, due to a strong internal magnetic field \citep{Osborne:2019iph}, accretion from a companion \citep{Ushomirsky:2000ax,Meadors:2015vpc,Haskell:2017ajb,Singh:2019dgy} or toroidal perturbations throughout the star, i.e. r-modes \citep{Owen:1998xg,mytidis2015constraining,mytidis2015sensitivity}. The size of these deformations, and the rate at which neutron stars accrete matter, are estimated to be small \citep{Lasky:2015uia}; thus, these processes would emit GWs at an almost fixed frequency by extracting rotational energy from the neutron star on a timescale much longer than the observation time of GW detectors~\citep{maggiore2008gravitational}. These are called continuous waves: quasi-monochromatic, long-duration GWs. Methods to search for CWs have been developed~\citep{riles2017recent,sieniawska2019continuous,Tenorio:2021wmz,Juliana:2022fbd} and are currently used for all-sky~\citep{KAGRA:2021una}, directed~\citep{LIGOScientific:2013wcb,Dergachev:2019pgs,piccinni2020directed,LIGOScientific:2021mwx}, targeted~\citep{LIGOScientific:2021hvc,LIGOScientific:2020lkw,LIGOScientific:2020gml,NarrowbandCW_LIGOScientific:2021quq} and post-merger remnant searches \citep{longpmr,abbott2017search,Oliver:2018dpt,Sun:2018hmm,PhysRevD.98.102004,miller2019effective,banagiri2019search}. CW methods have even been adapted to search for particle dark matter \citep{PierceRilesZhao2018,guo2019searching,Miller:2020vsl,LIGOScientific:2021odm}, boson clouds around black holes \citep{palomba2019direct,isi2019directed,Sun:2019mqb} and primordial black hole binaries \citep{Miller:2021knj,Miller:2020kmv}, all of which underscore the broad scope of CW physics. However, no method or search mentioned so far for neutron stars or dark matter, no matter how exotic, has allowed us to probe bulk properties of isolated neutron stars.

Though no CW has been detected yet, each type of CW search has shown promising results. Targeted searches continue to surpass the GW-amplitude spin-down limit, which assumes that all of the rotational energy lost by NSs as they spin-down is through GW radiation \citep{LIGOScientific:2021hvc,LIGOScientific:2020gml, LIGOScientific:2020lkw}. Additionally, all-sky and directed searches probe smaller and smaller deformations at galactic-centre distances \citep{LIGOScientific:2021inr,LIGOScientific:2022pjk}. The improved sensitivity of these searches over time brings us closer and closer to being able to make a detection of a CW from an isolated NS. Moreover, once we enter the detection era, GWs could be used as a novel messenger to identify new nearby NSs, 
an alternative to current searches for pulsars in electromagnetic (EM) data, whose average discovery rate is $\simeq 50\, \mathrm{yr^{-1}}$.
However, assuming the Galactic supernovae rate to be $10^{-2}\mathrm{yr^{-1}}$ \citep{Diehl:2006cf} and the age of the Milky Way to be $10^{10}$ yrs, there are roughly $\sim 10^{8}$ NSs \citep{Reed:2021scb} in our Galaxy alone. This large number of NSs implies that, even in the detection era, assuming a CW discovery rate of the same order or even ten times greater than the average electromagnetic one, we would still need centuries to individually detect the majority of the NSs and characterise the Galactic NS population properties. It is also not clear how many isolated NSs would need to be individually detected to make population-based statements. In addition, current CW searches are not yet designed to provide information on NS ensemble properties, even though methods have been developed to combine results from targeted searches for a few hundreds of known pulsars  \citep{CW_ensemble:2016PhRvD..94h4029F,Pitkin:PhysRevD.98.063001,buono2021method}. Nonetheless, this is just a small fraction of the larger population considered here, and
suggests the need for an alternative strategy to determine such properties that do not rely on measuring GWs from individual NSs.

We attempt to address this problem by searching for an astrophysical stochastic gravitational-wave background (SGWB)~\citep{2018PhRvL.120i1101A,2011PhRvD..84h4004R, 2011PhRvD..84l4037M, 2011ApJ73986Z, 2013MNRAS.431..882Z,2016PhRvD..94j3011D,Dhurandhar_Hotspot_PhysRevD.84.083007,HUGHES201486,2018arXiv180710620C,PhysRevD.50.1157,PhysRevD.85.023534,Regimbau_review,Inayoshi,Watanabe,Kamionkowski,Kosowsky,Turner} from the superposition of weak GW signals from individually-undetectable pulsars, which could already be observed by current detectors~\citep{Talukder:2014eba}. Its detection and characterisation would provide constraints that are independent and complementary to those inferred from CW (and EM) searches for individual NSs. Moreover, it would give insight into ensemble properties of NSs, by identifying certain traits (e.g. the mean value) of the statistical distributions of the parameters (e.g the ellipticity), which characterise the population of interest, at once.

In this work, we consider the Galactic-NS population and the NS populations of five "hotspots", i.e. patches of the sky that are expected to have a high number of NSs \citep{Dhurandhar_Hotspot_PhysRevD.84.083007,Okada_Hotspot_numerical_2012}, as potential candidates for our search. For each of these cases, we use cross-correlation \citep{allen-ottewill,Romano:2016dpx, Allen:1997ad} methods to search for a SGWB. Cross-correlation allows us to search for a common signal in multiple data streams simultaneously and disentangle it from instrumental noise. From these measurements, we could measure the number of Galactic NSs emitting in a given frequency band, and the average ellipticity of that population $\varepsilon_{\mathrm{av}}$. However, the search employed in this work finds no evidence of such a background; therefore, we set limits on the SGWB properties and then convert them into constraints on the average ellipticity of each population as a function of the number of NSs emitting GWs within the frequency band of the search.

The paper is organised as follows. In Section \ref{Sec:Model}, we describe the SGWB signal from isolated, rotating NSs, and model the NS population as a function of frequency using the known pulsars from the ATNF catalogue \citep{Manchester:2004bp}. In Section \ref{Sec:Methods}, we present the cross-correlation techniques and show how to estimate the average ellipticity of a population of pulsars from the results of our search for a SGWB signal. Then, in Section \ref{Sec:Results}, we illustrate the results of the searches for a SGWB from NSs when using the data from the first three observation runs of Advanced LIGO and Virgo over the population of Galactic NSs, and the NSs of the five hotspots: Virgo, Fornax, Antlia, Centaurus, and Hydra galaxies clusters, which contain thousands of galaxies and are assumed to have roughly 1000 times more NSs than our Galaxy. In addition to that, we express the results as limits on the average ellipticities of the considered populations. Finally, in Section \ref{Sec:Conclusions}, we summarise and discuss the implications of our results in terms of possible synergies between CW searches and SGWB ones, and future extensions to this work.

\section{Modelling the source}
\label{Sec:Model}
\begin{figure}
    \centering
    \includegraphics[width = \columnwidth]{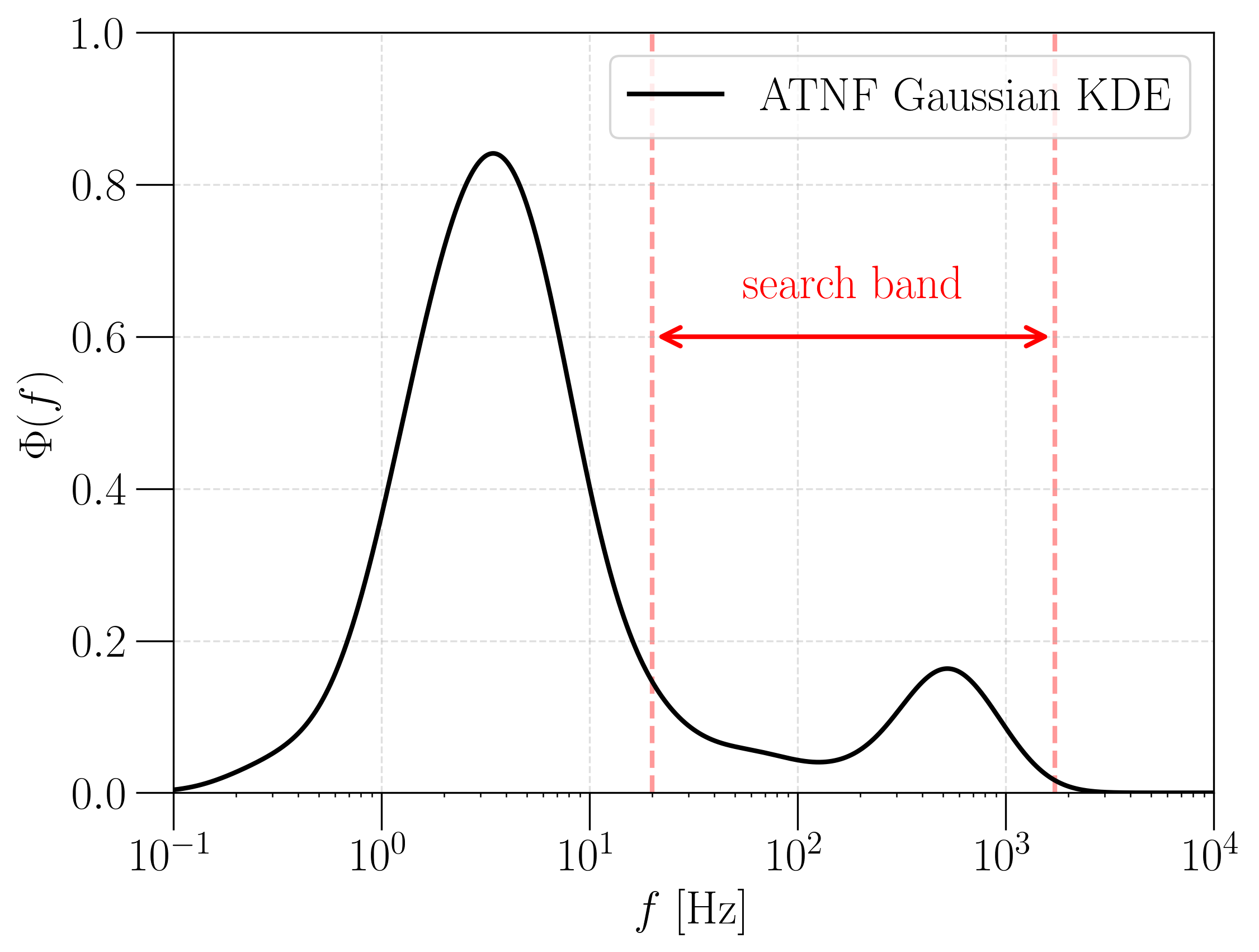}
    \caption{The frequency distribution of the PDF $\Phi(f)$, generated from the ATNF catalogue data. The red-dashed vertical lines show the frequency band 20-1726 Hz used in our search.}
    \label{fig:Phi(f)}
\end{figure}

\begin{figure}
    \centering
    \includegraphics[width = \columnwidth]{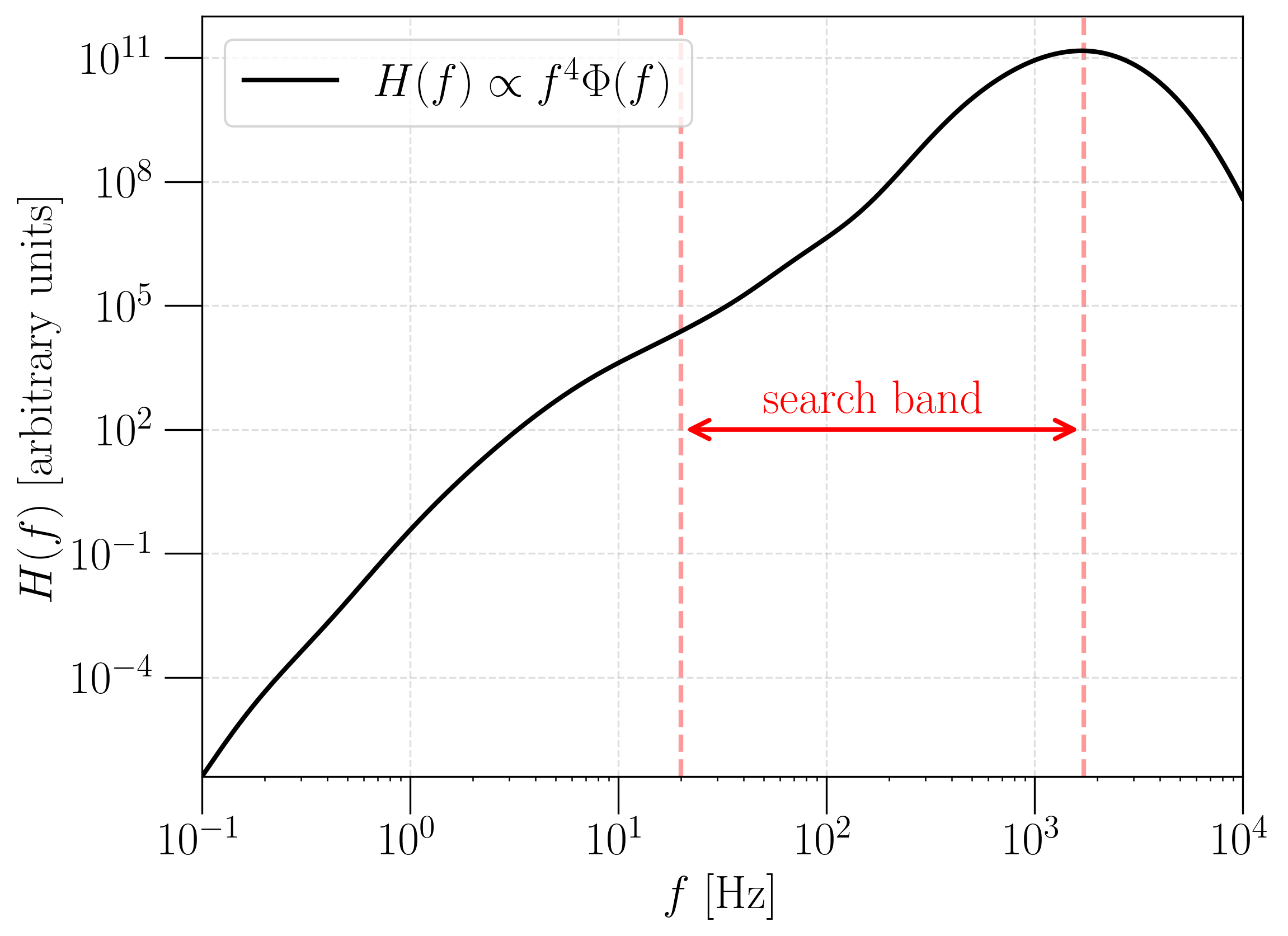}
    \caption{Unnormalised spectral shape $H(f)$ as a function of frequency, assuming $\Phi(f)$ derived from a Gaussian KDE using data from the ATNF catalogue. The red-dashed vertical lines show the frequency range that we analyse in our search.}
    \label{fig:H(f)_unscaled}
\end{figure}

The strain amplitude of a GW emitted from an isolated, rotating, non-axi-symmetric NS at a distance $d$ from Earth, with a moment of inertia along the z-axis $I_{zz}$, and an ellipticity $\varepsilon \equiv \frac{I_{xx} - I_{yy}}{I_{zz}}$, in the quadrupole approximation~\citep{maggiore2008gravitational}, is given as
\begin{equation}
\label{eq:h0(f)}
    h_{0}(f) = \frac{4 \upi^2 \, G \, \varepsilon \, I_{zz} }{c^4 \, d} \,f^2 \,,
\end{equation}
where $G$ is Newton's gravitational constant, $c$ is the speed of light, and $f$ is the frequency of the emitted GW, which is twice the rotational frequency of the NS.
Using equation \eqref{eq:h0(f)}, it is possible to show that (see appendix \ref{app:H(f)}) an ensemble of pulsars, whose contributions are summed incoherently, generates a GW power spectral density $H(f)$
\begin{equation}
\label{eq:spectral_shape}
    H(f) = \frac{32 \upi^4 G^2 \left\langle \varepsilon ^2 \right\rangle_{\mathrm{NS}} \left\langle I_{zz}^2 \right\rangle_{\mathrm{NS}}}{5 c^8}\, \left\langle \frac{1}{d^2} \right\rangle_{\mathrm{NS}}\, f^4 \, N(f) \,,
\end{equation}
where the angular brackets $\left\langle ... \right\rangle_{\mathrm{NS}}$ denote the ensemble average over the NS population, and $N(f)$ is the number of NSs emitting GWs between frequencies $f$ and $f+df$. To completely determine the frequency dependence of this stochastic signal, we rewrite $N(f)$ as
\begin{equation}
     N(f) = N_0 \, \Phi(f),
\end{equation}
where $N_0$ is the number of NSs in a given ensemble, and $\Phi(f)$ is the corresponding probability distribution function (PDF) of the NSs frequencies, defined such that
\begin{equation}
    N_0 \, \int_{-\infty}^{\infty} \, \Phi(f) \, df = N_0 \,.
\end{equation}
We consider $N_0\sim 10^{8}$ when studying the SGWB from Galactic NS, and $N_0 \sim 10^{11}$ when analysing the NS hotspots.
To model $\Phi(f)$, we employ an observation-driven approach~\citep{Reed:2021scb}, and use information about known pulsars available in the ATNF catalogue~\citep{Manchester:2004bp}. We start from the ($\log_{10}$-) frequency distribution of the $\simeq 3000$ pulsars available in the catalogue and obtain the frequency distribution PDF through a Gaussian kernel-density-estimator (KDE) \citep{2020SciPy-NMeth}. The resulting $\Phi(f)$ is shown in Figure \ref{fig:Phi(f)}. Even though the number of NSs used to construct $\Phi(f)$ is just a tiny fraction of the pulsars within our Galaxy and clusters of galaxies, the frequency distribution is expected to not be significantly biased by selection effects \citep{Lorimer:2008se, Lorimer:2019xjx} for millisecond pulsars. Moreover, this distribution is consistent with those obtained from population synthesis models~\citep{Story_2007, Talukder:2014eba}. 
From Figure \ref{fig:Phi(f)}, it is interesting to note that $\Phi(f)$ displays a secondary peak at $526$ Hz, which incidentally falls within the frequency band to which ground-based GW detectors are sensitive.
The unnormalised spectral shape $H(f)$ corresponding to the computed $\Phi(f)$ is shown in Figure \ref{fig:H(f)_unscaled}. 
In this spectrum, due to the dominant contribution from the $f^4$ term in equation \eqref{eq:spectral_shape}, the peak is shifted to a higher frequency ($1688$ Hz).

Given $\Phi(f)$ and $N_0$, we define $N_{\mathrm{band}}$ as the ``in-band'' NS number, which quantifies the number of NSs between a lower frequency $f_{\mathrm{min}}$ and a higher frequency $f_{\mathrm{max}}$. Thus, we can write $N_{\mathrm{band}}$ for the SGWB search described in this work as
\begin{equation}
\label{eq:Nband}
    N_{\mathrm{band}} = N_{0} \int^{f_{\mathrm{max}}}_{f_{\mathrm{min}}} \Phi(f) \, df = N_{0} \int^{1726 \mathrm{\, Hz}}_{20 \mathrm{\, Hz}} \Phi(f) \, df \simeq 0.16 \, N_0 \,,
\end{equation}
where the 20-1726 Hz denotes our chosen frequency band for this search\footnote{The frequency range used in this analysis is consistent with the previous stochastic searches \citep{O3_Iso_PhysRevD.104.022004, O3_Aniso_PhysRevD.104.022005}}. We analyse this frequency band because current GW detectors are most sensitive to signals arising between 20-1726 Hz \citep{Thrane_Romano_sensitivities_PhysRevD.88.124032}, and it contains the GW emission-band of millisecond pulsars. We also note that only $16\%$ of NSs emit GWs in our chosen frequency band. However, this fraction translates into $\sim 10^7$ and $\sim10^{10}$ in-band NSs in the galactic and hotspot cases, respectively, all emitting CWs, whose superposition would give rise to a continuous, Gaussian (due to the central limit theorem) SGWB. In this study, we rely on the spectral shape given in equation \eqref{eq:spectral_shape} to describe the SGWB from NS populations. 
By assuming a canonical value for $\left\langle I_{zz}^2 \right\rangle_{\mathrm{NS}}^{1/2} = 1.1\times 10^{38} \mathrm{kg \, m^2}$, and a fiducial value $\left\langle 1/d^2\right\rangle_{\mathrm{NS}}^{-1/2}$ for each population, we can translate the measurement of a SGWB into constraints on $\varepsilon_{\mathrm{av}}$ as a function of $N_{\mathrm{band}}$.
It is worth noting here that, along with equation \eqref{eq:spectral_shape}, it would be interesting to
consider the angular distribution of pulsars to characterise the stochastic GW signal. This is because, from current observations, the Galactic NS angular distribution is likely to be anisotropic, i.e. peaked towards the galactic plane \citep{lorimer_2012}, and the hotspots are localised in specific sky regions. However, in this work, we treat the SGWB from Galactic NSs as isotropic and consider the average power strain of the hotspots.
Considering all the anisotropies in the SGWB sky would require us to employ the matched-filtering ``$\lambda$-statistic'' proposed in~\cite{lambda_statistic}, and produce a template bank, which is out of the scope of the present work.

\section{Search methods}
\label{Sec:Methods}
Searches for a SGWB typically characterise the fractional energy density $\Omega_{\mathrm{GW}}$ \citep{Allen:1997ad, allen-ottewill,Cornish:2015pda, mingarelli2019understanding}, which is defined as the ratio between $\rho_{\mathrm{GW}}$, the energy density from all GWs in the Universe, and $\rho_{c} \equiv \frac{3 H_0^2 c^2}{8 \upi G}$, the critical density needed to have a flat Universe. Here, $H_0 = 67.9 \, \mathrm{km\, s^{-1} \, Mpc^{-1}}$ \citep{Planck:2015fie} is Hubble's parameter today.
$\Omega_{\mathrm{GW}}$ receives contributions from GWs at all frequencies and it is natural to study its frequency spectrum 
\begin{equation}
    \label{eq:omega_f}
    \Omega_{\mathrm{gw}}(f) = \frac{f}{\rho_{c}} \, \frac{d\rho_{\mathrm{gw}}(f)}{df} \,,
\end{equation}
which is related to $H(f)$ by
\begin{equation}
    \label{eq:omega_f_and_H_f}
\Omega_{\mathrm{gw}}(f) = \frac{2\upi^2}{3H_0^2} \, f^3 \, H(f) \,.
\end{equation}
Since we are performing a search that assumes GW sources to be isotropically distributed in the sky, equation \eqref{eq:omega_f} fully characterises the SGWB.

\subsection{The cross-correlation search for an isotropic SGWB}
\label{sec:iso_analysis}
As illustrated in the above discussion, we first perform a search for a Gaussian, stationary, unpolarised, isotropic SGWB. We use GW strain data from the LIGO-Hanford (H), LIGO-Livingston (L), and Virgo (V) detectors, and follow the search procedures in \cite{O3_Iso_PhysRevD.104.022004}. For each detector pair, called a ``baseline $IJ$'' ($I, J = {H, L, V}$), we divide the time-series output $s_I(t)$ in segments of duration $\tau$, labelled by $t$, take their Fourier-transforms $\tilde{s}_I(t;\,f)$, and calculate a cross-correlation statistic in each segment. Thus, we can define the cross-correlation estimator at every frequency, usually referred as the ``narrowband estimator'', as~\citep{Romano:2016dpx}
\begin{equation}
    \label{eq:CC_iso}
    \hat{C}_{IJ}(t; f) = \frac{2}{\tau} \, \frac{\Re[\tilde{s}_{I}^{*}(t; f) \,  \tilde{s}_{J}(t; f)]}{\gamma_{IJ}(f) \, S_{0}(f)},
\end{equation}
where $S_{0}(f) = (3H_0^2)/(10 \upi^2 f^3)$, and $\gamma_{IJ}(f)$ is the normalised isotropic overlap reduction function (ORF)~\citep{Allen:1997ad,Christensen_ORF_PhysRevD.46.5250,Flanagan_ORF_PhysRevD.48.2389} of the baseline $IJ$. The ORF quantifies the reduction in sensitivity due to the geometry of baseline and its response to the GW signal. 
The normalisation of the ORF is done in such a way that $\left\langle \hat{C}^{IJ}(t; f) \right\rangle = \Omega_{\mathrm{gw}}(f)$ in the absence of correlated noise. 
The variance associated with the above estimator, in the small signal limit, can be expressed as
\begin{equation}
    \label{sigma_CC_iso}
    \sigma^{2}_{IJ}(t; f) \approx \frac{1}{2\tau \,\Delta\, f}\frac{P_I(t; f)P_J(t; f)}{\gamma_{IJ}^2(f) \, S_0^2(t; f)} \,,
\end{equation}
where $P_I(t; f)$ is the one-sided power spectral density (PSD) in a detector, while $\Delta f$ denotes the frequency resolution.

Starting from the narrowband estimator $\hat{C}_{IJ}(t; f)$, we can build a broadband optimal estimator $\hat{C}_{IJ}$ by combining the cross-correlation spectra from different frequencies with appropriate weight factors. This optimal estimator and the associated uncertainty can be expressed as
\begin{align}
    \label{eq:broadband_CC_iso}
    \hat{C}_{IJ} &= \frac{\sum_{k, t} w(f_{\mathrm{k}}) \, \hat{C}_{IJ}(t;\,f_{\mathrm{k}}) \, \sigma^{-2}_{IJ}(t;\, f_{\mathrm{k}})}{\sum_{k} w^2(f_{\mathrm{k}})\sigma^{-2}_{IJ}(t;\, f_{\mathrm{k}})} \,, \\
    \sigma^{-2}_{IJ} &=  \sum_{k, t} w^2(f_{\mathrm{k}}) \, \sigma^{-2}_{IJ}(t; f_{\mathrm{k}}) \,,
\end{align}
where ${f_{\mathrm{k}}}$ is a set of discrete frequencies. The weights $w(f)$ can be derived for a generic $\Omega_{\mathrm{gw}}(f)$ following an optimal filtering approach \citep{Romano:2016dpx}
\begin{equation}
    \label{eq:weights_iso}
    w(f) = \frac{\Omega_{\mathrm{gw}}(f)}{\Omega_{\mathrm{gw}}(f_{\mathrm{ref}})} \,,
\end{equation}
where $f_{\mathrm{ref}}$ is an arbitrary reference frequency, fixed at $f_{\mathrm{ref}} = 25\; \mathrm{Hz}$ in this analysis to be consistent with \citep{O3_Iso_PhysRevD.104.022004}. After calculating the cross-correlation statistics for each pair of detectors, we can combine the individual broadband estimators from independent baselines (HL, HV, LV) as well as past observing runs (O1-HL, O2-HL) to obtain the final estimator $\hat{C}$ and its uncertainty:
\begin{align}
    \label{eq:CC_iso_final}
    &\hat{C} = \frac{\sum_{IJ} \hat{C}^{IJ} \, \sigma^{-2}_{IJ}}{\sum_{IJ} \sigma^{-2}_{IJ}}\\
    &\sigma^{-2} =  \sum_{IJ} \sigma^{-2}_{IJ} \,.
\end{align}
Now, we can recast $\Omega(f)$ in equation \eqref{eq:omega_f_and_H_f} (by means of equation \eqref{eq:spectral_shape})and rearranging the terms) in the following form:
\begin{equation}
    \label{eq:iso_assumed_omega}
    \Omega_{\mathrm{gw}}(f) = \Omega_{\mathrm{ref}} \left( \frac{f}{f_{\mathrm{ref}}} \right)^{7} \frac{\Phi(f)}{\Phi(f_{\mathrm{ref}})} \,.
\end{equation}
Finally, after applying the relevant data quality cuts \citep{O3_iso_data,O3_Iso_PhysRevD.104.022004}, and vetoing all outliers found in our search, we can use the estimators presented in equation \eqref{eq:CC_iso_final} to set upper limits on $\Omega_{\mathrm{ref}}$ through a Bayesian analysis for the model of interest. To do that, we employ the likelihood
\begin{equation}
    \label{eq:lklhood_CC}
    p\left( \hat{C}(f_{\mathrm{k}}) | \Omega(f_{\mathrm{k}}) \right) = \frac{1}{\sqrt{2\upi}\sigma(f_{\mathrm{k}})} \,
    \exp{\left[-\frac{\left(\hat{C}(f_{\mathrm{k}}) - \Omega(f_{\mathrm{k}})\right)^2}{2\sigma^2(f_{\mathrm{k}})}\right]} \,,
\end{equation}
where $\hat{C}(f_{\mathrm{k}})$ is assumed to be Gaussian distributed in absence of a signal \citep{Romano:2016dpx} and $\Omega(f_{\mathrm{k}})$ is the model for the SGWB in equation \eqref{eq:iso_assumed_omega}.
Moreover, we can also use the estimator for $\Omega_{\mathrm{ref}}$ as a starting point to obtain a constraint on $\varepsilon_{\mathrm{av}}$ of a NS population as a function of $N_{\mathrm{band}}$, which will be discussed next.

\subsection{Constraining the ellipticity of a NS population}
\label{sec:ellipticity_population}

Here, we show how to translate the results of the above-presented analysis to build an estimator for the average ellipticity of a NS population. We recall from section \ref{Sec:Model} that we are using fiducial values for $\left\langle I_{zz}^2 \right\rangle_{\mathrm{NS}}^{1/2}$ and $\left\langle 1/d^2\right\rangle_{\mathrm{NS}}^{-1/2}$, while $\varepsilon_{\mathrm{av}}$ and $N_{\mathrm{band}}$ are left as free parameters for the time being.

Considering equation \eqref{eq:spectral_shape}, along with the frequency range of interest, we can rewrite equation \eqref{eq:omega_f_and_H_f} as
\begin{equation}
    \label{eq:omega_vs_shape}
    \Omega(f) = 
    \frac{64 \upi^6 G^2}{3 H_0^2} \frac{\left\langle \varepsilon ^2 \right\rangle_{\mathrm{NS}} \left\langle I_{zz}^2 \right\rangle_{\mathrm{NS}}}{5 c^8}\, \left\langle \frac{1}{d^2} \right\rangle_{\mathrm{NS}}\, f^7 \, N_{\mathrm{band}} \Phi(f) \,.
\end{equation}
Then, by combining the above equation with equations \ref{eq:weights_iso} and \ref{eq:iso_assumed_omega}, we obtain
\begin{equation}
    \label{eq:omega_vs_ellipticity}
    \Omega(f) = \left( \frac{f}{f_{\mathrm{ref}}} \right)^{7} \frac{\Phi(f)}{\Phi(f_{\mathrm{ref}})} \, \xi \left\langle \varepsilon^2 \right\rangle_{\mathrm{NS}} = w(f) \, \xi \left\langle \varepsilon^2 \right\rangle_{\mathrm{NS}}\, ,
\end{equation}
where we have introduced $\xi = \xi(N_{\mathrm{band}}) \equiv \Omega_{\mathrm{ref}}/\left\langle \varepsilon^2 \right\rangle_{\mathrm{NS}}$, which is just a proportionality constant, once $N_{\mathrm{band}}$ is fixed.
Within this framework, using equation \eqref{eq:CC_iso}, the above equation can be recast in terms of different narrowband estimators:
\begin{equation}
\label{eq:eps_hat_om_hat}
    \left(\widehat{\varepsilon^2}\right)_{\mathrm{av}}(f_{\mathrm{k}}) = \frac{1}{\xi}\frac{\hat{C}_{IJ}(f_{\mathrm{k}})}{w(f_{\mathrm{k}})} \equiv \frac{\hat{\Omega}_{\mathrm{ref}}(f_{\mathrm{k}})}{\xi}\,,
\end{equation}
where $\hat{\Omega}_{\mathrm{ref}}(f_{\mathrm{k}})$ is the narrowband estimator of $\Omega_{\mathrm{ref}}$, while $ \left(\widehat{\varepsilon^2}\right)_{\mathrm{av}}(f_{\mathrm{k}})$ is the narrowband estimator of the average squared ellipticity $\left\langle \varepsilon^2 \right\rangle_{\mathrm{NS}}$ of the NS population\footnote{The frequencies $f_{\mathrm{k}}$ in equation \eqref{eq:eps_hat_om_hat} must be interpreted as labels and not as functional dependence.}. 

Starting from the above estimator, we can derive the relationship between $ \left(\widehat{\varepsilon^2}\right)_{\mathrm{av}}(f_{\mathrm{k}})$ and the average ellipticity of the NS population along with its estimator.
This can be done by writing the expectation value of $ \left(\widehat{\varepsilon^2}\right)_{\mathrm{av}}(f_{\mathrm{k}})$:
\begin{equation}
    \label{eq:epsilon2_expectation}
    \left\langle  \left(\widehat{\varepsilon^2}\right)_{\mathrm{av}}(f_{\mathrm{k}}) \right\rangle = \left\langle \varepsilon^2(f_{\mathrm{k}}) \right\rangle_{\mathrm{NS}} \equiv  \varepsilon^2_{\mathrm{av}}(f_{\mathrm{k}}) + \sigma_{\varepsilon}^2 (f_{\mathrm{k}})\,,
\end{equation}
where $\varepsilon_{\mathrm{av}}(f_{\mathrm{k}}) \equiv \left\langle \varepsilon(f_{\mathrm{k}}) \right\rangle_{\mathrm{NS}}$ is the mean value of the ellipticity, while $\sigma_{\varepsilon}^2 (f_{\mathrm{k}})$ is the intrinsic variance of the ellipticity distribution. 
Then, from equation \eqref{eq:epsilon2_expectation}, we can define the biased estimator of the average ellipticity
\begin{equation}
    \label{eq:ellipticity_estimator}
    \hat{\varepsilon}_{\mathrm{av}}(f_{\mathrm{k}}) \equiv \sqrt{ \left(\widehat{\varepsilon^2}\right)_{\mathrm{av}}(f_{\mathrm{k}})} \,.
\end{equation}
The bias introduced from the non-zero variance of the ellipticity distribution should be small, since the physical ellipticity is a positive definite quantity. Thus, one can assume $\sigma_{\varepsilon} (f_{\mathrm{k}}) \lesssim \varepsilon_{\mathrm{av}}(f_{\mathrm{k}})$ and ignore the variance in equation \eqref{eq:epsilon2_expectation}. 
This choice translates into more conservative constraints derived from $\hat{\varepsilon} (f_{\mathrm{k}})$\footnote{A non-zero $\sigma_{\varepsilon}^2 (f_{\mathrm{k}})$ will increase the intensity of a stochastic signal at a fixed $\varepsilon_{\mathrm{av}} (f_{\mathrm{k}})$, making its detection easier.}. Possible ways to account for the bias, as in the case of a detection of a SGWB from a NS population, would be to estimate $\sigma_{\varepsilon}^2 (f_{\mathrm{k}})$ from the measurements of individual NSs (such as the ones detected with CW/EM techniques) from theoretical models of the population.

Given the estimator $\hat{\varepsilon}_{\mathrm{av}}(f_{\mathrm{k}})$, we can derive the associated uncertainty $\sigma_{\hat{\varepsilon}}(f_{\mathrm{k}})$ from the likelihood function $p_{\varepsilon} \left(\hat{\varepsilon}_{\mathrm{av}}(f_{\mathrm{k}})|\varepsilon_{\mathrm{av}}(f_{\mathrm{k}})\right)$. 
To obtain this likelihood, we use equations \eqref{eq:omega_vs_ellipticity}, \eqref{eq:eps_hat_om_hat}, and \eqref{eq:ellipticity_estimator} to express $\Omega(f_\mathrm{k})$ and $\hat{C}(f_\mathrm{k})$ as a function of $\varepsilon_{\mathrm{av}}(f_\mathrm{k})$ and $\hat{\varepsilon}_{\mathrm{av}}(f_\mathrm{k})$, and we perform a change of variables in equation \eqref{eq:lklhood_CC}. Following this prescription, we obtain the following likelihood function for $\hat{\varepsilon}_{\mathrm{av}}(f_\mathrm{k})$, which is no longer a Gaussian:
\begin{equation}
    \label{eq:lklhood_epsilon}
    p_{\varepsilon} \left(\hat{\varepsilon}_{\mathrm{av}}(f_{\mathrm{k}})|\varepsilon_{\mathrm{av}}(f_{\mathrm{k}})\right) = \sqrt{\frac{8}{\upi}} \frac{\varepsilon_{\mathrm{av}}(f_{\mathrm{k}}) \, \xi}{ \sigma_{\hat{\Omega}}(f_{\mathrm{k}})} \, \exp{\left[-\frac{(\hat{\varepsilon}^{2}_{\mathrm{av}}(f_{\mathrm{k}}) - \varepsilon^2_{\mathrm{av}}(f_{\mathrm{k}}))^2 \, \xi^2}{2\sigma^2_{\hat{\Omega}}(f_{\mathrm{k}})}\right]} \,,
\end{equation}
where $\sigma_{\hat{\Omega}}(f_\mathrm{k})$ is the error corresponding to $\hat{\Omega}_{\mathrm{ref}}(f_\mathrm{k})$. By applying the definition of variance to the above distribution, in the limit $\hat{\varepsilon}_{\mathrm{av}}(f_{\mathrm{k}}) << 1$, we arrive at:
\begin{equation}
    \label{eq:sigma_epsilon}
    \sigma^2_{\hat{\varepsilon}}(f_{\mathrm{k}})|_{\hat{\varepsilon} << 1} \approx \left[ \sqrt{\frac{2}{\upi}} - \frac{2^{3/2}\upi}{\Gamma^2(\frac{1}{4})} \right] \frac{\sigma_{\hat{\Omega}}(f_{\mathrm{k}})}{\xi}  \simeq 0.12 \frac{\sigma_{\hat{\Omega}}(f_{\mathrm{k}})}{\xi} \,. 
\end{equation}
The derivation and the expression of $\sigma^2_{\hat{\varepsilon}}(f_{\mathrm{k}})$ in the general case $\hat{\varepsilon}_{\mathrm{av}}(f_{\mathrm{k}})>0$ are reported in Appendix \ref{app:sigma_eps}.

Finally, assuming the ellipticity to be independent of the frequency, the narrowband estimators $\hat{\varepsilon}_{\mathrm{av}}(f_{\mathrm{k}})$ can be combined to obtain the optimal broadband estimator $\hat{\varepsilon}_{\mathrm{opt}}$, with a relative uncertainty $\sigma_{\mathrm{opt}}$ as
\begin{equation}
    \label{eq:epsilon_hat_opt}
    \hat{\varepsilon}_{\mathrm{opt}} = \frac{\sum_{k} \hat{\varepsilon}_{\mathrm{av}}(f_{\mathrm{k}})\sigma^{-2}_{\hat{\varepsilon}}(f_{\mathrm{k}})}{\sum_{k} \sigma^{-2}_{\hat{\varepsilon}}(f_{\mathrm{k}})}, \qquad \sigma_{\mathrm{opt}} = \left( \sum_{k} \sigma^{-2}_{\hat{\varepsilon}}(f_{\mathrm{k}}) \right)^{-1/2} \,.
\end{equation}
Using equation \eqref{eq:omega_vs_shape} and plugging it into the above equation, the optimal estimator depends on the number of in-band NSs, which has been considered as a free parameter in the analysis, through the relation $\hat{\varepsilon}_{\mathrm{opt}} \propto N_{\mathrm{band}}^{-1/2}$. Hence, the upper limits on the average ellipticity $\varepsilon_{\mathrm{av}}$ will also depend on $N_{\mathrm{band}}$. In this situation, we could set upper limits on quantities such as $\varepsilon_{\mathrm{av}} N_{\mathrm{band}}^{1/2}$ or $\varepsilon^2_{\mathrm{av}} N_{\mathrm{band}}$, which are inherently independent from $N_{\mathrm{band}}$. Alternatively, we could evaluate upper limits on $\varepsilon_{\mathrm{av}}$ at a reference value of $N_{\mathrm{band}}$, and then map them into the $N_{\mathrm{band}}-\varepsilon_{\mathrm{av}}$ plane. In this paper, we follow the second approach to present constraints on the average ellipticity.

\begin{figure}
    \includegraphics[width = \columnwidth]{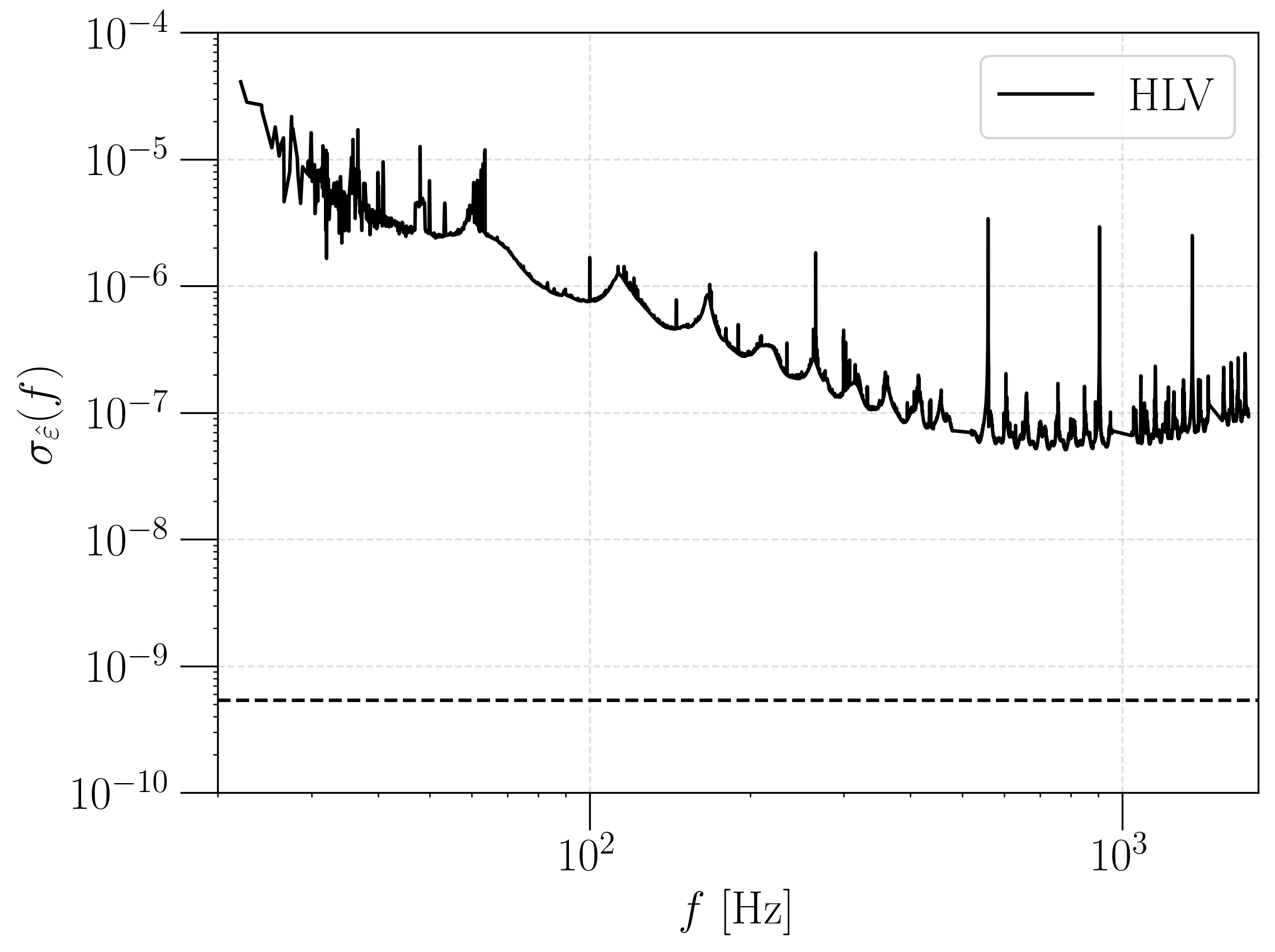}
    \caption{Plot of the $1\sigma$ sensitivity to the average ellipticity of Galactic NSs. The solid curve shows the uncertainty $\sigma_{\varepsilon}(f_{\mathrm{k}})$ associated to the narrowband estimators, while the dashed one is the broadband value of $\sigma_{\mathrm{opt}}$. The improvement of the search sensitivity by combining the narrowband estimators ranges between two and four orders of magnitude. The plot assumes $N_{\mathrm{band}}= 1.6 \times 10^{7}$, $\left\langle 1/d^2\right\rangle_{\mathrm{NS}}^{-1/2} = 6$ kpc, and $\left\langle I_{zz}^2 \right\rangle_{\mathrm{NS}}^{1/2} = 10^{38}\, \mathrm{kg\,  m^2}$.}
    \label{fig:1-sigma-epsilon}
\end{figure}

\begin{figure}
    \includegraphics[width = \columnwidth]{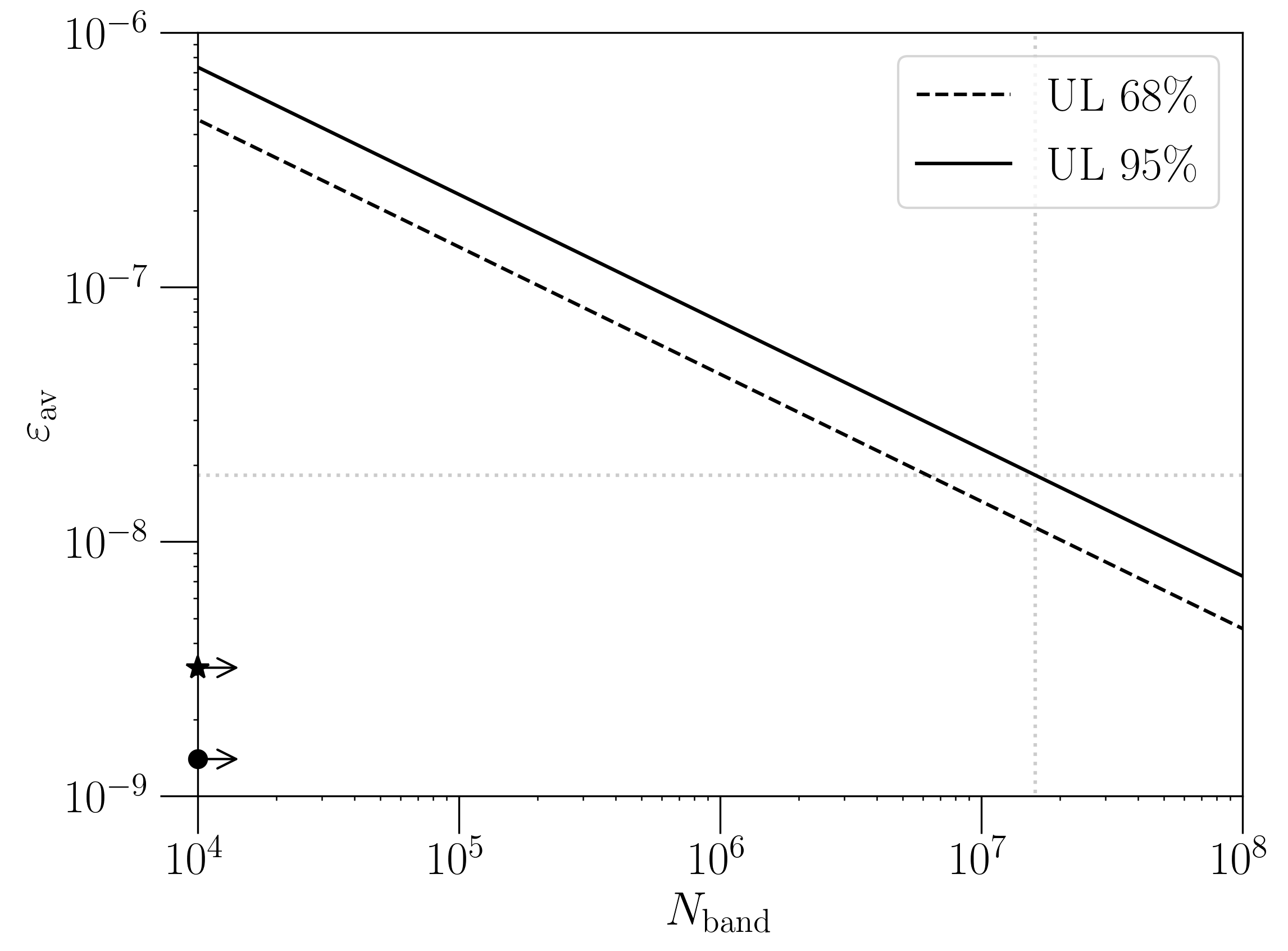}
    \caption{68\% (dashed) and 95\% (solid) confidence-level Bayesian upper limits in the $N_{\mathrm{band}}-\varepsilon_{\mathrm{av}}$ plane, assuming a log-uniform prior on $\varepsilon_{\mathrm{av}}$. Here, we have set $N_{\mathrm{band}}$ to range from $10^{4}$ and $10^{8}$. The dotted grey lines identify the 95\% upper limit on $\varepsilon_{\mathrm{av}}$ obtained with the pivot value of in-band NSs, $N_{\mathrm{band}} = 1.6 \times 10^7$. The star and the circle on the y-axis denote the most recent, lowest upper-limits on a single NS ellipticity (\textit{independent of $N_{\mathrm{band}}$}), respectively $\varepsilon \lesssim 3.2 \times 10^{-9}$ from targeted \citep{LIGOScientific:2021hvc} CW searches and $\varepsilon \lesssim 1.4 \times 10^{-9}$ from all-sky \citep{LIGOScientific:2022pjk} ones.}
    \label{fig:ULs}
\end{figure}

\section{Results of the analyses}
\label{Sec:Results}
We perform this analysis on publicly available data~\citep{gwosc:LIGOScientific:2019lzm} from the first three observing runs (O1, O2, and O3) of the Advanced LIGO and Virgo detectors. We first search for an isotropic SGWB from Galactic NSs, assuming the model given in equation \eqref{eq:iso_assumed_omega}. 
Using these search results, we also place upper limits on the average ellipticity of the NS population. Secondly, we consider five clusters of galaxies as hotspots of GWs. By associating a specific patch\footnote{The methodology to build the sky patches and evaluate the associated average background are presented in appendix \ref{app:BBR}, and makes use of the radiometer search \citep{Ballmer_2006,MitraML:PhysRevD.77.042002,radiometer}.} in the sky to each of them, we again set constraints on the average ellipticity of the NS population starting from the results for the fractional SGWB energy density for each sky patch analysed, $\Omega^{\mathrm{patch}}_{\mathrm{gw}}(f)$. These results are given in the following subsections.

\subsection{Galactic NS results}
\label{sec:iso_results}
Since our search for an isotropic SGWB from Galactic isolated NSs did not find any evidence for a signal, we set upper limits on $\Omega_{\mathrm{ref}}$. These results are subsequently used to constrain $\varepsilon_{\mathrm{av}}$, and are listed in table~\ref{tab:iso_omega}. The first four columns from the left contain the results for the SGWB background search. The second column shows the value of the cross-correlation statistic and the associated 1$\sigma$ uncertainty derived from equation \eqref{eq:CC_iso_final}. The third and fourth columns present the 95\% confidence-level Bayesian upper limits for $\Omega_{\mathrm{ref}}$. These upper limits are obtained by marginalising the likelihood function given in equation \eqref{eq:lklhood_CC} over a uniform (third column) and a log-uniform prior (fourth column) on the strength of the SGWB. It is worth noting that the log-uniform prior seems to be the most natural choice, since $\Omega_{\mathrm{ref}}$ range is expected to span several orders of magnitude, and is more sensitive to small signals. The log-uniform prior range was chosen to be between $10^{-18} \leq \Omega_{\mathrm{ref}} \leq 10^{-8}$. The upper bound is large enough such that there is no posterior support at that value, while the lower bound cannot be zero for this kind of prior. On the other hand, we have also included the result from the uniform prior case, which leads to more conservative upper limits. In both cases, the estimator for $\Omega_{\mathrm{ref}}$, as well as the upper limits, are all of $\mathcal{O}(10^{-14})$. These values are significantly smaller than those for other power-law models for $\Omega_{\mathrm{gw}}(f)$, e.g. those reported in \cite{O3_Iso_PhysRevD.104.022004}, since the $\Omega_{\rm gw}$ used in this paper is dominated by the $\sim f^7$ term.

The last two columns in table~\ref{tab:iso_omega} illustrate the limits we have obtained on the average ellipticity at $1\sigma$ sensitivity and the corresponding $95\%$ Bayesian upper limits on $\varepsilon_{\mathrm{av}}$, using the fiducial value $\left\langle 1/d^2\right\rangle^{-1/2}= 6\; \mathrm{kpc}$ in equation \eqref{eq:spectral_shape}. The value of the estimator $\hat{\varepsilon}_{\mathrm{opt}}$ is of $\mathcal{O}(10^{-11})$, with an associated uncertainty one order of magnitude larger.
The improvement in the sensitivity of the search, that comes from combining the estimators over frequencies, is illustrated in figure \ref{fig:1-sigma-epsilon}. Here, the relative uncertainties associated to $\hat{\varepsilon}_{\mathrm{av}}(f_{\mathrm{k}})$ and $\hat{\varepsilon}_{\mathrm{opt}}$ are plotted as a function of frequency.

On the other hand, the Bayesian upper limit on the average ellipticity $\varepsilon_{\mathrm{av}}$ has been obtained using the likelihood function in equation \eqref{eq:lklhood_epsilon} by assuming a log-uniform prior in the range $10^{-12}-10^{-4}$. The obtained constraint is of $\mathcal{O}(10^{-8})$. As discussed in section \ref{sec:ellipticity_population}, the constraint holds only for the representative value of $N_{\mathrm{band}}$ that we have explicitly presented here. However, it can be easily mapped into the $N_{\mathrm{band}}-\varepsilon_{\mathrm{av}}$ plane for different values of in-band NSs. Considering the range $N_{\mathrm{band}} \in [10^{4}-10^{8}]$, we present the 68\% and 95\% Bayesian upper limits on the average ellipticity in figure \ref{fig:ULs}. In this figure, the pivot value $N_{\mathrm{band}} = 1.6 \times 10^{7}$ is highlighted using a dotted line for an easy comparison. We also report the latest, lowest upper limits on a NS ellipticity from targeted \citep{LIGOScientific:2021hvc} and all-sky \citep{LIGOScientific:2022pjk} CW searches on the y-axis.
It is evident from the figure that the resulting $\varepsilon_{\mathrm{av}}$, ranging between $10^{-8}$ and $10^{-6}$, follows the $\varepsilon_{\mathrm{av}} \propto N_{\mathrm{band}}^{-1/2}$ relation, as anticipated in section \ref{sec:ellipticity_population}.
\begin{table*}
    \centering
    \begin{tabular}{ |c|c|c|c|c|c|c|c| }
        \hline
        $\Omega(f)$ & $\hat{C}^{\mathrm{O1+O2+O3}}/(10^{-14})$  & $\Omega_{\mathrm{ref}}^{95\%,\, \mathrm{Uniform}}$ & $\Omega_{\mathrm{ref}}^{95\%,\, \mathrm{Log-uniform}}$ & $\Phi(f)$ & $N_{\mathrm{band}}$ & $\hat{\varepsilon}_{\mathrm{opt}}^{\mathrm{O1+O2+O3}}/10^{-11}$ & $\varepsilon^{95\%}_{\mathrm{Log-uniform}}$\\
        \hline
        $\propto \left( f \right)^7 \Phi(f)$ & $0.9 \pm 1.9$ & $4.5\times 10^{-14}$ & $2.0\times 10^{-14}$ & ATNF-KDE  & $1.6 \times 10^{7}$ & $2.5 \pm 53.5$ & $1.8\times 10^{-8}$\\ 
        \hline
    \end{tabular}
    \caption{Results of the isotropic search for a SGWB from an ensemble of Galactic NSs using data from the first three LIGO-Virgo-KAGRA observing runs, and the subsequent constraints on the average ellipticity of the Galactic NS population. The first four columns are the results from our search, in which $\Omega(f)$, the cross-correlation statistics, and the upper limits on $\Omega_{\mathrm{ref}}$, using a uniform and log-uniform prior, are reported. The last four columns encode information about the Galactic NS population, such as $\Phi(f)$ and $N_{\mathrm{band}}$, the average ellipticity optimal estimator, and the upper limit obtained by assuming a log-uniform prior on $\varepsilon$ between $10^{-12}-10^{-4}$.}
    \label{tab:iso_omega}
\end{table*}

\subsection{Hotspot results}
\label{sec:hotspot_results}
\begin{figure}
    \includegraphics[width=\columnwidth]{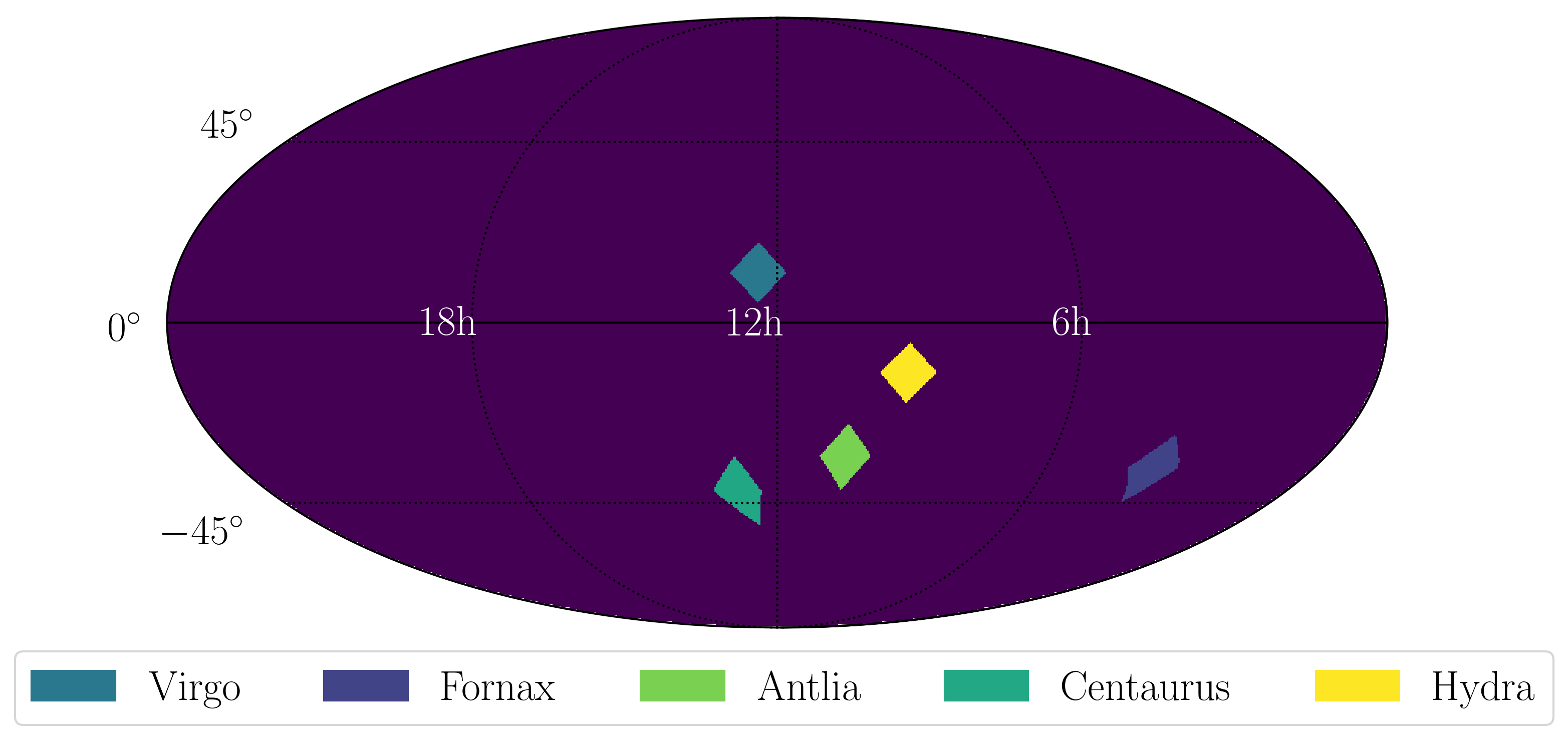}
    \caption{The sky-patches associated with the five NS hotspots: Virgo, Fornax, Antlia, Centaurus, and Hydra clusters. Each patch consists of 9 pixels with $N_{\mathrm{side}}= 16$: the central one being the one associated with the galaxy cluster, and the eight closest neighbours. The sky map is represented as a Mollweide projection of the sky in ecliptic coordinates.}
    \label{fig:sky-patches}
\end{figure}

The search results from five NS GW hotspots and the corresponding constraints on their ellipticities are reported in table \ref{tab:patches_ellipticity}. First, we pixelate the sky by employing the {\tt{HEALPix}} (Hierarchical Equal Area isoLatitude Pixelation) pixelisation scheme \citep{HEALPix:Gorski:2004by, Healpy:Zonca2019}, with $N_{\mathrm{side}}=16$ (3072 pixels, each one with an extension of $\simeq 13.4 \, \mathrm{deg}^2$). From the right ascension and declination of the hotspot, we identify one pixel and its eight closest neighbours. These collections of pixels will act as a patch in the pixelated sky and are illustrated in figure~\ref{fig:sky-patches}. The signal model for each of the hotspots is similar to the one used in the Galactic NS analysis, except for the number of in-band NS (which in this case is $N_{\mathrm{band}} = 1.6 \times 10^{10}$), and the distance parameter $\left\langle 1/d^2\right\rangle_{\mathrm{NS}}^{-1/2}$ (values considered are shown in the second column of table \ref{tab:patches_ellipticity}). For each hotspot, we first estimated the $\Omega^{\mathrm{patch}}_{\mathrm{gw}}(f)$ using the folded data~\citep{Folding:PhysRevD.92.022003,ASAF:LIGOScientific:2021oez} and {\tt PyStoch} pipeline~\citep{PyStoch:Ain:2018zvo}, and then followed the method described in section \ref{sec:ellipticity_population} to derive constraints on the average ellipticity of the NS populations in each of the hotspots. 

Within the above framework, we have derived the optimal estimators and the 95\% confidence upper limits related to the average ellipticity of the NS populations of the hotspots. These quantities are respectively reported in the third and fourth columns of table \ref{tab:patches_ellipticity}. 
Because of the absence of any detection, we set upper limits
using the same criteria as in the Galactic case.
We find that the estimators are of the order of $10^{-10}-10^{-9}$, whereas the upper limits of the average ellipticity are around $10^{-7}-10^{-6}$. Comparing the constraints and the relative hotspots distances, we note that the constraints become less stringent when the source is more distant from Earth.
By contrasting the hotspot limits with those from Galactic NSs, we observe that the former are one or two orders of magnitude larger than the latter. This difference could arise from several factors, from the model assumed to the characteristics of the ground-based detectors.
From a modelling perspective, based on equations \ref{eq:spectral_shape} and \ref{eq:omega_vs_shape}, the difference between these two kinds of NS populations is encoded in the average of the inverse squared distance of the source from the Earth, the number of in-band NS, and the size of the examined region of the sky. The hotspot populations are estimated to have $10^{3}$ times more NSs compared to the Galactic population but are also $10^3$ times more distant from Earth. This means that in light of the model considered and the assumed values, the average ellipticity of the cluster NS population should be $\simeq \sqrt{10^{3}}$ times the Galactic one, in the naive case where the two SGWB have the same intensity. From the detector perspective, instead, the intrinsic sensitivity of the instrument to the source distance and its position in the sky have an impact. The detector becomes less sensitive the more distant the source of interest is. The sensitivity may get even worse if the source is well localised and spends most of the time in the region of the sky where the detectors have poorly observed modes \citep{Romano:2016dpx}.
Given two populations with different positions and spreads in the sky, analyses of these two areas may lead to less stringent upper limits of one population with respect to another one, even though the properties of the two populations' original signals are the same. This effect can be mitigated by combining data from multiple detectors (pairs) of a detector network, but cannot be completely suppressed, due to the intrinsic geometry of the network and its interaction with the GW signal.
Further studies to evaluate the impact of the model and choice and the detector network characteristics in the recovery of the signal will be the subject of future work.

\begin{table}
    \begin{tabular}{ |c|c|c|c|c| }
        \hline
        Hotspot & $\left\langle 1/d^2 \right\rangle_{\mathrm{NS}}^{-1/2}$ (Mpc) & $\hat{\varepsilon}^{\mathrm{O1+O2+O3}}_{\mathrm{opt}}/10^{-9}$ & $\varepsilon^{95\%}_{\mathrm{Log-uniform}}/10^{-7}$ \\
        \hline
        Virgo & 18 & $0.6 \pm 10.6 $ & $3.6$ \\
        Fornax & 19 & $0.5 \pm 10.1 $ & $3.5$ \\
        Antlia & 40.7 & $1.5 \pm 22.1 $ & $7.6$  \\
        Centaurus & 52.4 & $1.4 \pm 27.9$ & $9.6$ \\
        Hydra  & 58.3 & $3.8 \pm 34.2 $ & $11.8$ \\
        \hline
    \end{tabular}
    \caption{Relevant parameters and results of searches for NSs in hotspots. For each cluster of galaxies, a fiducial value of $\left\langle 1/d^2 \right\rangle_{\mathrm{NS}}^{-1/2}$ (second column), the broadband estimator $\hat{\varepsilon}_{\mathrm{opt}}$ (third column), and the 95\% confidence level Bayesian upper limits on the average ellipticity of the population (fourth column) are reported. The upper limits have been obtained by assuming a log-uniform prior between $10^{-12}-10^{-4}$ over the ellipticities.}
    \label{tab:patches_ellipticity}
\end{table}

\section{Discussions and conclusions}
\label{Sec:Conclusions}
In this work, we have derived constraints on the average ellipticity of a NS population from the results of a cross-correlation-based search for a SGWB. We have considered two classes of NS populations: those in our Galaxy and those in five extra-galactic clusters, which we call NS hotspots. We have not found compelling evidence of a SGWB signal from any of the considered sources and hence have set upper limits on the intensity of the background by bounding the energy density parameter $\Omega_{\mathrm{gw}}(f)$. These results have then been translated to constraints of the NS average ellipticity, obtained to be as low as $\varepsilon_{\mathrm{av}} \lesssim 1.8 \times 10^{-8}$ with $N_{\mathrm{band}}=1.6\times 10^{7}$ in the case of Galactic NS and $\varepsilon_{\mathrm{av}} \la [3.5-11.8]\times 10^{-7}$ with $N_{\mathrm{band}}=1.6\times 10^{10}$ for those in galaxy clusters. These constraints obtained using the data from the first three observation runs of Advanced LIGO and Virgo are the first of their kind.

If we consider recent results from CW searches for Galactic NSs, whose lowest limits on NSs ellipticities are $\varepsilon \lesssim 3.2 \times 10^{-9}$ (for J0636+5129) from targeted searches \citep{LIGOScientific:2021hvc} and $\varepsilon \lesssim 1.4 \times 10^{-9}$ (for a NS at 10 pc from Earth and at $2047.5$ Hz) from all-sky searches \citep{LIGOScientific:2022pjk}, we observe that they are one order of magnitude lower than the values reported here. It is not straightforward to compare these limits, since these analyses constrain different properties of NSs. Targeted CW searches are more sensitive to individual NS properties, such as the ellipticity, but must obtain their limits only based on known pulsars. Moreover, all-sky searches have proven to be computationally expensive (order of months to run), and can only search in certain parts of the parameter space semi-coherently, which limits their sensitivities to ellipticities of $\mathcal{O}(10^{-7})$ and $\mathcal{O}(10^{-5})$ at $\mathcal{O}(1)$ kpc at high and low frequencies, respectively \citep{LIGOScientific:2022pjk}. Instead, searches for SGWB have become computationally efficient and faster (order of days to run) \citep{O3_Iso_PhysRevD.104.022004,O3_Aniso_PhysRevD.104.022005, Folding:PhysRevD.92.022003,PyStoch:Ain:2018zvo,pystochSpH}, but their constraining power is weaker compared to targeted CW searches. In addition to that, they have the advantage (once the results are available) of instantaneously identifying the features of an ensemble of known or unknown NSs, which would otherwise require decades/centuries to be determined through individual NS discoveries. Because SGWB and CW searches attempt to answer different physical questions, they can work in synergy. Using the methods of the former, it would be possible to perform rapid, blind all-sky searches for NS signals and transmit the coordinates of possible outliers as inputs of the latter, for a more refined and sensitive search.

In this work, we have restricted ourselves to constrain the average ellipticity of a NS population, given the number of in-band NSs. We have assumed values for the average squared moment of inertia and the average square inverse distance of the population. We could gain even more information about NS populations by treating these quantities as free parameters. Additionally, we could estimate and set constraints on these quantities through a full Bayesian search, in which priors could be derived from population synthesis simulations. These simulations could also be used to model the NS frequency and angular distributions, which could then be used as an alternative to those derived from the ATNF catalogue, especially in the Extra-Galactic case. 
Moreover, the inclusion of angular distribution of the NSs would allow to perform a template-based matched-filtering search using the $\lambda-$ statistics from \cite{lambda_statistic}, which may set less conservative upper limits.
Finally, from the synthesised population, the corresponding SGWB signal could be simulated, and its prospects for detection and characterisation could be examined within the networks of the future detector.
Two ways of doing this would be to consider a network, where KAGRA and the future LIGO-India \citep{LIGO_India:Saleem_2021} are included, or considering the next-generation interferometers, such as Einstein Telescope \citep{ET:Punturo:2010zz} and Cosmic Explorer \citep{CE:Reitze:2019iox}, and evaluate their impact on these kinds of searches.
These possibilities will be explored in future work. 

\section*{Acknowledgements}
This material is based upon work supported by NSF's LIGO Laboratory which is a major facility fully funded by the National Science Foundation. This research has used data obtained from the Gravitational Wave Open Science Center, a service of LIGO Laboratory, the LIGO Scientific Collaboration and the Virgo Collaboration. LIGO Laboratory and Advanced LIGO are funded by the United States National Science Foundation (NSF) as well as the Science and Technology Facilities Council (STFC) of the United Kingdom, the Max-Planck-Society (MPS), and the State of Niedersachsen/Germany for support of the construction of Advanced LIGO and construction and operation of the GEO600 detector. Additional support for Advanced LIGO was provided by the Australian Research Council. Virgo is funded, through the European Gravitational Observatory (EGO), by the French Centre National de Recherche Scientifique (CNRS), the Italian Istituto Nazionale della Fisica Nucleare (INFN) and the Dutch Nikhef, with contributions by institutions from Belgium, Germany, Greece, Hungary, Ireland, Japan, Monaco, Poland, Portugal, and Spain.

We acknowledge the use of Caltech LDAS clusters and the supercomputing facilities of the Universit\'e catholique de Louvain (CISM/UCL) and the Consortium des \'Equipements de Calcul Intensif en F\'ed\'eration Wallonie Bruxelles (C\'ECI), funded by the Fond de la Recherche Scientifique de Belgique (F.R.S.-FNRS) under convention 2.5020.11 and by the Walloon Region. The authors gratefully acknowledge the support of the NSF, STFC, INFN and CNRS for provision of computational resources.

The authors thank Patrick Meyers for carefully reading the manuscript and providing valuable comments. This work significantly benefited from the interactions with the Stochastic Working Group of the LIGO-Virgo-KAGRA Scientific Collaboration. We also thank T. Regimbau, D. Agarwal, Brendan T. Reed, Alex Deibel, and C. J. Horowitz for useful discussions. 

F.D.L. is supported by a FRIA Grant of the Belgian Fund for Research, F.R.S.-FNRS. A.L.M. is a beneficiary of a FSR Incoming Postdoctoral Fellowship. This article has a LIGO document number LIGO-P2200050. 

\section*{Data Availability}
This research has made use of data or software obtained from the Gravitational Wave Open Science Center (gw-openscience.org) \citep{gwosc:LIGOScientific:2019lzm}, a service of LIGO Laboratory, the LIGO Scientific Collaboration, the Virgo Collaboration, and KAGRA.
Publicly available ``Data for Upper Limits on the Isotropic Gravitational-Wave Background from Advanced LIGO's and Advanced Virgo's Third Observing Run'' from \cite{O3_iso_data} has been used as well within the stochastic analysis.

\bibliographystyle{mnras}
\bibliography{bibliography} 

\begin{thebibliography}{}
\makeatletter
\relax
\def\mn@urlcharsother{\let\do\@makeother \do\$\do\&\do\#\do\^\do\_\do\%\do\~}
\def\mn@doi{\begingroup\mn@urlcharsother \@ifnextchar [ {\mn@doi@}
  {\mn@doi@[]}}
\def\mn@doi@[#1]#2{\def\@tempa{#1}\ifx\@tempa\@empty \href
  {http://dx.doi.org/#2} {doi:#2}\else \href {http://dx.doi.org/#2} {#1}\fi
  \endgroup}
\def\mn@eprint#1#2{\mn@eprint@#1:#2::\@nil}
\def\mn@eprint@arXiv#1{\href {http://arxiv.org/abs/#1} {{\tt arXiv:#1}}}
\def\mn@eprint@dblp#1{\href {http://dblp.uni-trier.de/rec/bibtex/#1.xml}
  {dblp:#1}}
\def\mn@eprint@#1:#2:#3:#4\@nil{\def\@tempa {#1}\def\@tempb {#2}\def\@tempc
  {#3}\ifx \@tempc \@empty \let \@tempc \@tempb \let \@tempb \@tempa \fi \ifx
  \@tempb \@empty \def\@tempb {arXiv}\fi \@ifundefined
  {mn@eprint@\@tempb}{\@tempb:\@tempc}{\expandafter \expandafter \csname
  mn@eprint@\@tempb\endcsname \expandafter{\@tempc}}}

\bibitem[\protect\citeauthoryear{Aasi et~al.}{Aasi
  et~al.}{2013}]{LIGOScientific:2013wcb}
Aasi J.,  et~al., 2013, \mn@doi [Phys. Rev. D] {10.1103/PhysRevD.88.102002},
  88, 102002

\bibitem[\protect\citeauthoryear{{Aasi} et~al.,}{{Aasi}
  et~al.}{2015}]{2015CQGra..32g4001L}
{Aasi} J.,  et~al., 2015, \mn@doi [CQGra] {10.1088/0264-9381/32/7/074001},
  \href {http://adsabs.harvard.edu/abs/2015CQGra..32g4001L} {32, 074001}

\bibitem[\protect\citeauthoryear{Abbott}{Abbott}{2021a}]{O3_Iso_PhysRevD.104.022004}
Abbott R. e.~a.,  2021a, \mn@doi [Phys. Rev. D] {10.1103/PhysRevD.104.022004},
  104, 022004

\bibitem[\protect\citeauthoryear{Abbott}{Abbott}{2021b}]{O3_Aniso_PhysRevD.104.022005}
Abbott R. e.~a.,  2021b, \mn@doi [Phys. Rev. D] {10.1103/PhysRevD.104.022005},
  104, 022005

\bibitem[\protect\citeauthoryear{{Abbott} et~al.}{{Abbott}
  et~al.}{2007}]{radiometer}
{Abbott} B.,  et~al., 2007, \mn@doi [\prd] {10.1103/PhysRevD.76.082003}, \href
  {http://adsabs.harvard.edu/abs/2007PhRvD..76h2003A} {76, 082003}

\bibitem[\protect\citeauthoryear{Abbott et~al.}{Abbott
  et~al.}{2017}]{abbott2017search}
Abbott B.~P.,  et~al., 2017, \mn@doi [Astrophys. J. Lett.]
  {10.3847/2041-8213/aa9a35}, 851, L16

\bibitem[\protect\citeauthoryear{{Abbott} et~al.,}{{Abbott}
  et~al.}{2018}]{2018PhRvL.120i1101A}
{Abbott} B.~P.,  et~al., 2018, \mn@doi [Physical Review Letters]
  {10.1103/PhysRevLett.120.091101}, \href
  {http://adsabs.harvard.edu/abs/2018PhRvL.120i1101A} {120, 091101}

\bibitem[\protect\citeauthoryear{Abbott et~al.}{Abbott et~al.}{2019}]{longpmr}
Abbott B.~P.,  et~al., 2019, \mn@doi [The Astrophysical Journal]
  {10.3847/1538-4357/ab0f3d}, 875, 160

\bibitem[\protect\citeauthoryear{Abbott et~al.}{Abbott
  et~al.}{2020}]{LIGOScientific:2020gml}
Abbott R.,  et~al., 2020, \mn@doi [Astrophys. J. Lett.]
  {10.3847/2041-8213/abb655}, 902, L21

\bibitem[\protect\citeauthoryear{Abbott et~al.}{Abbott
  et~al.}{2021f}]{ASAF:LIGOScientific:2021oez}
Abbott R.,  et~al., 2021f, {All-sky, all-frequency directional search for
  persistent gravitational-waves from Advanced LIGO's and Advanced Virgo's
  first three observing runs} (\mn@eprint {arXiv} {2110.09834})

\bibitem[\protect\citeauthoryear{Abbott et~al.,}{Abbott
  et~al.}{2021c}]{LIGOScientific:2021odm}
Abbott R.,  et~al., 2021c, Constraints on dark photon dark matter using data
  from LIGO's and Virgo's third observing run (\mn@eprint {arXiv} {2105.13085})

\bibitem[\protect\citeauthoryear{Abbott et~al.}{Abbott
  et~al.}{2021e}]{O3_iso_data}
Abbott R.,  et~al., 2021e, {Data for Upper Limits on the Isotropic
  Gravitational-Wave Background from Advanced LIGO's and Advanced Virgo's Third
  Observing Run}, \url {{https://dcc.ligo.org/LIGO-G2001287/public}}

\bibitem[\protect\citeauthoryear{Abbott et~al.}{Abbott
  et~al.}{2021b}]{NarrowbandCW_LIGOScientific:2021quq}
Abbott R.,  et~al., 2021b, {Narrowband searches for continuous and
  long-duration transient gravitational waves from known pulsars in the
  LIGO-Virgo third observing run} (\mn@eprint {arXiv} {2112.10990})

\bibitem[\protect\citeauthoryear{Abbott et~al.}{Abbott
  et~al.}{2021d}]{LIGOScientific:2021inr}
Abbott R.,  et~al., 2021d, {Search of the Early O3 LIGO Data for Continuous
  Gravitational Waves from the Cassiopeia A and Vela Jr. Supernova Remnants}
  (\mn@eprint {arXiv} {2111.15116})

\bibitem[\protect\citeauthoryear{Abbott et~al.}{Abbott
  et~al.}{2021a}]{LIGOScientific:2021hvc}
Abbott R.,  et~al., 2021a, {Searches for Gravitational Waves from Known Pulsars
  at Two Harmonics in the Second and Third LIGO-Virgo Observing Runs}
  (\mn@eprint {arXiv} {2111.13106})

\bibitem[\protect\citeauthoryear{Abbott et~al.}{Abbott
  et~al.}{2021g}]{gwosc:LIGOScientific:2019lzm}
Abbott R.,  et~al., 2021g, \mn@doi [SoftwareX] {10.1016/j.softx.2021.100658},
  13, 100658

\bibitem[\protect\citeauthoryear{Abbott et~al.}{Abbott
  et~al.}{2021h}]{KAGRA:2021una}
Abbott R.,  et~al., 2021h, \mn@doi [Phys. Rev. D]
  {10.1103/PhysRevD.104.082004}, 104, 082004

\bibitem[\protect\citeauthoryear{Abbott et~al.}{Abbott
  et~al.}{2021i}]{LIGOScientific:2020lkw}
Abbott R.,  et~al., 2021i, \mn@doi [Astrophys. J.] {10.3847/2041-8213/abffcd},
  913, L27

\bibitem[\protect\citeauthoryear{Abbott et~al.}{Abbott
  et~al.}{2021j}]{LIGOScientific:2021mwx}
Abbott R.,  et~al., 2021j, \mn@doi [Astrophys. J.] {10.3847/1538-4357/ac17ea},
  921, 80

\bibitem[\protect\citeauthoryear{Abbott et~al.}{Abbott
  et~al.}{2022}]{LIGOScientific:2022pjk}
Abbott R.,  et~al., 2022, All-sky search for continuous gravitational waves
  from isolated neutron stars using Advanced LIGO and Advanced Virgo O3 data
  (\mn@eprint {arXiv} {"2201.00697"})

\bibitem[\protect\citeauthoryear{{Acernese} et~al.,}{{Acernese}
  et~al.}{2015}]{2015CQGra..32b4001A}
{Acernese} F.,  et~al., 2015, \mn@doi [CQGra] {10.1088/0264-9381/32/2/024001},
  \href {http://adsabs.harvard.edu/abs/2015CQGra..32b4001A} {32, 024001}

\bibitem[\protect\citeauthoryear{Ade et~al.}{Ade et~al.}{2016}]{Planck:2015fie}
Ade P. A.~R.,  et~al., 2016, \mn@doi [Astron. Astrophys.]
  {10.1051/0004-6361/201525830}, 594, A13

\bibitem[\protect\citeauthoryear{Ain, Dalvi  \& Mitra}{Ain
  et~al.}{2015}]{Folding:PhysRevD.92.022003}
Ain A.,  Dalvi P.,   Mitra S.,  2015, \mn@doi [Phys. Rev. D]
  {10.1103/PhysRevD.92.022003}, 92, 022003

\bibitem[\protect\citeauthoryear{Ain, Suresh  \& Mitra}{Ain
  et~al.}{2018}]{PyStoch:Ain:2018zvo}
Ain A.,  Suresh J.,   Mitra S.,  2018, \mn@doi [Phys. Rev. D]
  {10.1103/PhysRevD.98.024001}, 98, 024001

\bibitem[\protect\citeauthoryear{Akutsu et~al.}{Akutsu
  et~al.}{2021}]{KAGRA:2020tym}
Akutsu T.,  et~al., 2021, \mn@doi [PTEP] {10.1093/ptep/ptaa125}, 2021, 05A101

\bibitem[\protect\citeauthoryear{{Allen} \& {Ottewill}}{{Allen} \&
  {Ottewill}}{1997}]{allen-ottewill}
{Allen} B.,  {Ottewill} A.~C.,  1997, \mn@doi [\prd] {10.1103/PhysRevD.56.545},
  \href {http://adsabs.harvard.edu/abs/1997PhRvD..56..545A} {56, 545}

\bibitem[\protect\citeauthoryear{Allen \& Romano}{Allen \&
  Romano}{1999}]{Allen:1997ad}
Allen B.,  Romano J.~D.,  1999, \mn@doi [Phys. Rev. D]
  {10.1103/PhysRevD.59.102001}, 59, 102001

\bibitem[\protect\citeauthoryear{Ballmer}{Ballmer}{2006}]{Ballmer_2006}
Ballmer S.~W.,  2006, \mn@doi [Classical and Quantum Gravity]
  {10.1088/0264-9381/23/8/s23}, 23, S179

\bibitem[\protect\citeauthoryear{Banagiri, Sun, Coughlin  \& Melatos}{Banagiri
  et~al.}{2019}]{banagiri2019search}
Banagiri S.,  Sun L.,  Coughlin M.~W.,   Melatos A.,  2019, \mn@doi [Phys. Rev.
  D] {10.1103/PhysRevD.100.024034}, 100, 024034

\bibitem[\protect\citeauthoryear{Bar-Kana}{Bar-Kana}{1994}]{PhysRevD.50.1157}
Bar-Kana R.,  1994, \mn@doi [Phys. Rev. D] {10.1103/PhysRevD.50.1157}, 50, 1157

\bibitem[\protect\citeauthoryear{Buono, Rosa, D'Onofrio, Errico, Palomba,
  Piccinni  \& Sequino}{Buono et~al.}{2021}]{buono2021method}
Buono M.,  Rosa R.~D.,  D'Onofrio L.,  Errico L.,  Palomba C.,  Piccinni O.~J.,
    Sequino V.,  2021, \mn@doi [Classical and Quantum Gravity]
  {10.1088/1361-6382/abf1c0}, 38, 135021

\bibitem[\protect\citeauthoryear{Christensen}{Christensen}{1992}]{Christensen_ORF_PhysRevD.46.5250}
Christensen N.,  1992, \mn@doi [Phys. Rev. D] {10.1103/PhysRevD.46.5250}, 46,
  5250

\bibitem[\protect\citeauthoryear{Cook \& Sorbo}{Cook \&
  Sorbo}{2012}]{PhysRevD.85.023534}
Cook J.~L.,  Sorbo L.,  2012, \mn@doi [Phys. Rev. D]
  {10.1103/PhysRevD.85.023534}, 85, 023534

\bibitem[\protect\citeauthoryear{Cornish \& Romano}{Cornish \&
  Romano}{2015}]{Cornish:2015pda}
Cornish N.~J.,  Romano J.~D.,  2015, \mn@doi [Phys. Rev. D]
  {10.1103/PhysRevD.92.042001}, 92, 042001

\bibitem[\protect\citeauthoryear{Cusin, Pitrou  \& Uzan}{Cusin
  et~al.}{2017}]{PhysRevD.96.103019}
Cusin G.,  Pitrou C.,   Uzan J.-P.,  2017, \mn@doi [Phys. Rev. D]
  {10.1103/PhysRevD.96.103019}, 96, 103019

\bibitem[\protect\citeauthoryear{Cusin, Dvorkin, Pitrou  \& Uzan}{Cusin
  et~al.}{2018}]{PhysRevLett.120.231101}
Cusin G.,  Dvorkin I.,  Pitrou C.,   Uzan J.-P.,  2018, \mn@doi [Phys. Rev.
  Lett.] {10.1103/PhysRevLett.120.231101}, 120, 231101

\bibitem[\protect\citeauthoryear{Cusin, Durrer  \& Ferreira}{Cusin
  et~al.}{2019}]{2018arXiv180710620C}
Cusin G.,  Durrer R.,   Ferreira P.~G.,  2019, \mn@doi [Phys. Rev. D]
  {10.1103/PhysRevD.99.023534}, 99, 023534

\bibitem[\protect\citeauthoryear{Dergachev, Papa, Steltner  \&
  Eggenstein}{Dergachev et~al.}{2019}]{Dergachev:2019pgs}
Dergachev V.,  Papa M.~A.,  Steltner B.,   Eggenstein H.-B.,  2019, \mn@doi
  [Phys. Rev. D] {10.1103/PhysRevD.99.084048}, 99, 084048

\bibitem[\protect\citeauthoryear{Dhurandhar, Tagoshi, Okada, Kanda  \&
  Takahashi}{Dhurandhar et~al.}{2011}]{Dhurandhar_Hotspot_PhysRevD.84.083007}
Dhurandhar S.,  Tagoshi H.,  Okada Y.,  Kanda N.,   Takahashi H.,  2011,
  \mn@doi [Phys. Rev. D] {10.1103/PhysRevD.84.083007}, 84, 083007

\bibitem[\protect\citeauthoryear{Diehl et~al.}{Diehl
  et~al.}{2006}]{Diehl:2006cf}
Diehl R.,  et~al., 2006, \mn@doi [Nature] {10.1038/nature04364}, 439, 45

\bibitem[\protect\citeauthoryear{{Dvorkin}, {Uzan}, {Vangioni}  \&
  {Silk}}{{Dvorkin} et~al.}{2016}]{2016PhRvD..94j3011D}
{Dvorkin} I.,  {Uzan} J.-P.,  {Vangioni} E.,   {Silk} J.,  2016, \mn@doi [Phys.
  Rev. D] {10.1103/PhysRevD.94.103011}, \href
  {https://ui.adsabs.harvard.edu/abs/2016PhRvD..94j3011D} {94, 103011}

\bibitem[\protect\citeauthoryear{{Fan}, {Chen}  \& {Messenger}}{{Fan}
  et~al.}{2016}]{CW_ensemble:2016PhRvD..94h4029F}
{Fan} X.,  {Chen} Y.,   {Messenger} C.,  2016, \mn@doi [\prd]
  {10.1103/PhysRevD.94.084029}, \href
  {https://ui.adsabs.harvard.edu/abs/2016PhRvD..94h4029F} {94, 084029}

\bibitem[\protect\citeauthoryear{Flanagan}{Flanagan}{1993}]{Flanagan_ORF_PhysRevD.48.2389}
Flanagan E.~E.,  1993, \mn@doi [Phys. Rev. D] {10.1103/PhysRevD.48.2389}, 48,
  2389

\bibitem[\protect\citeauthoryear{G\'orski, Hivon, Banday, Wandelt, Hansen,
  Reinecke  \& Bartelman}{G\'orski et~al.}{2005}]{HEALPix:Gorski:2004by}
G\'orski K.~M.,  Hivon E.,  Banday A.~J.,  Wandelt B.~D.,  Hansen F.~K.,
  Reinecke M.,   Bartelman M.,  2005, \mn@doi [Astrophys. J.] {10.1086/427976},
  622, 759

\bibitem[\protect\citeauthoryear{Guo, Riles, Yang  \& Zhao}{Guo
  et~al.}{2019}]{guo2019searching}
Guo H.-K.,  Riles K.,  Yang F.-W.,   Zhao Y.,  2019, Communications Physics, 2,
  1

\bibitem[\protect\citeauthoryear{Haskell \& Patruno}{Haskell \&
  Patruno}{2017}]{Haskell:2017ajb}
Haskell B.,  Patruno A.,  2017, \mn@doi [Phys. Rev. Lett.]
  {10.1103/PhysRevLett.119.161103}, 119, 161103

\bibitem[\protect\citeauthoryear{Hughes}{Hughes}{2014}]{HUGHES201486}
Hughes S.~A.,  2014, \mn@doi [Physics of the Dark Universe]
  {https://doi.org/10.1016/j.dark.2014.10.003}, 4, 86

\bibitem[\protect\citeauthoryear{Inayoshi, Kashiyama, Visbal  \&
  Haiman}{Inayoshi et~al.}{2021}]{Inayoshi}
Inayoshi K.,  Kashiyama K.,  Visbal E.,   Haiman Z.,  2021, \mn@doi [Astrophys.
  J.] {10.3847/1538-4357/ac106d}, 919, 41

\bibitem[\protect\citeauthoryear{Isi, Sun, Brito  \& Melatos}{Isi
  et~al.}{2019}]{isi2019directed}
Isi M.,  Sun L.,  Brito R.,   Melatos A.,  2019, \mn@doi [Phys. Rev. D]
  {10.1103/PhysRevD.99.084042}, 99, 084042

\bibitem[\protect\citeauthoryear{Jenkins \& Sakellariadou}{Jenkins \&
  Sakellariadou}{2018}]{PhysRevD.98.063509}
Jenkins A.~C.,  Sakellariadou M.,  2018, \mn@doi [Phys. Rev. D]
  {10.1103/PhysRevD.98.063509}, 98, 063509

\bibitem[\protect\citeauthoryear{{Jenkins}, {Sakellariadou}, {Regimbau}  \&
  {Slezak}}{{Jenkins} et~al.}{2018}]{2018PhRvD..98f3501J}
{Jenkins} A.~C.,  {Sakellariadou} M.,  {Regimbau} T.,   {Slezak} E.,  2018,
  \mn@doi [Phys. Rev. D] {10.1103/PhysRevD.98.063501}, \href
  {https://ui.adsabs.harvard.edu/abs/2018PhRvD..98f3501J} {98, 063501}

\bibitem[\protect\citeauthoryear{Jones \& Andersson}{Jones \&
  Andersson}{2002}]{Jones:2001yg}
Jones D.~I.,  Andersson N.,  2002, \mn@doi [Mon. Not. Roy. Astron. Soc.]
  {10.1046/j.1365-8711.2002.05180.x}, 331, 203

\bibitem[\protect\citeauthoryear{Juliana}{Juliana}{2022}]{Juliana:2022fbd}
Juliana P.~O.,  2022, Status and perspectives of Continuous Gravitational Wave
  searches (\mn@eprint {arXiv} {"2202.01088"})

\bibitem[\protect\citeauthoryear{Kamionkowski, Kosowsky  \&
  Turner}{Kamionkowski et~al.}{1994}]{Kamionkowski}
Kamionkowski M.,  Kosowsky A.,   Turner M.~S.,  1994, \mn@doi [Phys. Rev. D]
  {10.1103/PhysRevD.49.2837}, 49, 2837

\bibitem[\protect\citeauthoryear{Kosowsky, Turner  \& Watkins}{Kosowsky
  et~al.}{1992}]{Kosowsky}
Kosowsky A.,  Turner M.~S.,   Watkins R.,  1992, \mn@doi [Phys. Rev. D]
  {10.1103/PhysRevD.45.4514}, 45, 4514

\bibitem[\protect\citeauthoryear{Lasky}{Lasky}{2015}]{Lasky:2015uia}
Lasky P.~D.,  2015, \mn@doi [Publ. Astron. Soc. Austral.]
  {10.1017/pasa.2015.35}, 32, e034

\bibitem[\protect\citeauthoryear{{Lasky}, {Bennett}  \& {Melatos}}{{Lasky}
  et~al.}{2013}]{2013PhRvD..87f3004L}
{Lasky} P.~D.,  {Bennett} M.~F.,   {Melatos} A.,  2013, \mn@doi [Phys. Rev. D]
  {10.1103/PhysRevD.87.063004}, \href
  {https://ui.adsabs.harvard.edu/abs/2013PhRvD..87f3004L} {87, 063004}

\bibitem[\protect\citeauthoryear{Lorimer}{Lorimer}{2008}]{Lorimer:2008se}
Lorimer D.~R.,  2008, \mn@doi [Living Rev. Rel.] {10.12942/lrr-2008-8}, 11, 8

\bibitem[\protect\citeauthoryear{Lorimer}{Lorimer}{2012}]{lorimer_2012}
Lorimer D.~R.,  2012, \mn@doi [Proceedings of the International Astronomical
  Union] {10.1017/S1743921312023769}, 8, 237–242

\bibitem[\protect\citeauthoryear{Lorimer et~al.}{Lorimer
  et~al.}{2019}]{Lorimer:2019xjx}
Lorimer D.~R.,  et~al., 2019, {Radio Pulsar Populations} (\mn@eprint {arXiv}
  {1903.06526})

\bibitem[\protect\citeauthoryear{Maggiore}{Maggiore}{2007}]{maggiore2008gravitational}
Maggiore M.,  2007, {Gravitational Waves. Vol. 1: Theory and Experiments}

\bibitem[\protect\citeauthoryear{Manchester, Hobbs, Teoh  \& Hobbs}{Manchester
  et~al.}{2005}]{Manchester:2004bp}
Manchester R.~N.,  Hobbs G.~B.,  Teoh A.,   Hobbs M.,  2005, \mn@doi [Astron.
  J.] {10.1086/428488}, 129, 1993

\bibitem[\protect\citeauthoryear{{Marassi}, {Schneider}, {Corvino}, {Ferrari}
  \& {Portegies Zwart}}{{Marassi} et~al.}{2011}]{2011PhRvD..84l4037M}
{Marassi} S.,  {Schneider} R.,  {Corvino} G.,  {Ferrari} V.,   {Portegies
  Zwart} S.,  2011, \mn@doi [Phys. Rev. D] {10.1103/PhysRevD.84.124037}, \href
  {https://ui.adsabs.harvard.edu/abs/2011PhRvD..84l4037M} {84, 124037}

\bibitem[\protect\citeauthoryear{{Mazumder}, {Mitra}  \&
  {Dhurandhar}}{{Mazumder} et~al.}{2014}]{2014PhRvD..89h4076M}
{Mazumder} N.,  {Mitra} S.,   {Dhurandhar} S.,  2014, \mn@doi [Phys. Rev. D]
  {10.1103/PhysRevD.89.084076}, \href
  {http://adsabs.harvard.edu/abs/2014PhRvD..89h4076M} {89, 084076}

\bibitem[\protect\citeauthoryear{Meadors, Goetz  \& Riles}{Meadors
  et~al.}{2016}]{Meadors:2015vpc}
Meadors G.~D.,  Goetz E.,   Riles K.,  2016, \mn@doi [Class. Quant. Grav.]
  {10.1088/0264-9381/33/10/105017}, 33, 105017

\bibitem[\protect\citeauthoryear{Miller, Astone  et~al.}{Miller
  et~al.}{2018}]{PhysRevD.98.102004}
Miller A.,  Astone P.,   et~al., 2018, \mn@doi [Phys. Rev. D]
  {10.1103/PhysRevD.98.102004}, 98, 102004

\bibitem[\protect\citeauthoryear{Miller, Astone  et~al.}{Miller
  et~al.}{2019}]{miller2019effective}
Miller A.~L.,  Astone P.,   et~al., 2019, \mn@doi [Phys. Rev. D]
  {10.1103/PhysRevD.100.062005}, 100, 062005

\bibitem[\protect\citeauthoryear{Miller, Aggarwal, Clesse  \& De~Lillo}{Miller
  et~al.}{2021a}]{Miller:2021knj}
Miller A.~L.,  Aggarwal N.,  Clesse S.,   De~Lillo F.,  2021a, Constraints on
  planetary and asteroid-mass primordial black holes from continuous
  gravitational-wave searches (\mn@eprint {arXiv} {"2110.06188"})

\bibitem[\protect\citeauthoryear{Miller, Clesse, De~Lillo, Bruno, Depasse  \&
  Tanasijczuk}{Miller et~al.}{2021b}]{Miller:2020kmv}
Miller A.~L.,  Clesse S.,  De~Lillo F.,  Bruno G.,  Depasse A.,   Tanasijczuk
  A.,  2021b, \mn@doi [Phys. Dark Univ.] {10.1016/j.dark.2021.100836}, 32,
  100836

\bibitem[\protect\citeauthoryear{Miller et~al.}{Miller
  et~al.}{2021c}]{Miller:2020vsl}
Miller A.~L.,  et~al., 2021c, \mn@doi [Phys. Rev. D]
  {10.1103/PhysRevD.103.103002}, 103, 103002

\bibitem[\protect\citeauthoryear{Mingarelli, Taylor, Sathyaprakash  \&
  Farr}{Mingarelli et~al.}{2019}]{mingarelli2019understanding}
Mingarelli C. M.~F.,  Taylor S.~R.,  Sathyaprakash B.~S.,   Farr W.~M.,  2019,
  Understanding $\Omega_\mathrm{gw}(f)$ in Gravitational Wave Experiments
  (\mn@eprint {arXiv} {1911.09745})

\bibitem[\protect\citeauthoryear{Mitra, Dhurandhar, Souradeep, Lazzarini,
  Mandic, Bose  \& Ballmer}{Mitra et~al.}{2008}]{MitraML:PhysRevD.77.042002}
Mitra S.,  Dhurandhar S.,  Souradeep T.,  Lazzarini A.,  Mandic V.,  Bose S.,
  Ballmer S.,  2008, \mn@doi [Phys. Rev. D] {10.1103/PhysRevD.77.042002}, 77,
  042002

\bibitem[\protect\citeauthoryear{Mytidis, Coughlin  \& Whiting}{Mytidis
  et~al.}{2015}]{mytidis2015constraining}
Mytidis A.,  Coughlin M.,   Whiting B.,  2015, \mn@doi [The Astrophysical
  Journal] {10.1088/0004-637x/810/1/27}, 810, 27

\bibitem[\protect\citeauthoryear{Mytidis, Panagopoulos, Panagopoulos, Miller
  \& Whiting}{Mytidis et~al.}{2019}]{mytidis2015sensitivity}
Mytidis A.,  Panagopoulos A.~A.,  Panagopoulos O.~P.,  Miller A.,   Whiting B.,
   2019, \mn@doi [Phys. Rev. D] {10.1103/PhysRevD.99.024024}, 99, 024024

\bibitem[\protect\citeauthoryear{Okada, Kanda, Dhurandhar, Tagoshi  \&
  Takahashi}{Okada et~al.}{2012}]{Okada_Hotspot_numerical_2012}
Okada Y.,  Kanda N.,  Dhurandhar S.,  Tagoshi H.,   Takahashi H.,  2012,
  \mn@doi [J. Phys. Conf. Ser.] {10.1088/1742-6596/363/1/012040}, 363, 012040

\bibitem[\protect\citeauthoryear{Oliver, Keitel  \& Sintes}{Oliver
  et~al.}{2019}]{Oliver:2018dpt}
Oliver M.,  Keitel D.,   Sintes A.~M.,  2019, \mn@doi [Phys. Rev. D]
  {10.1103/PhysRevD.99.104067}, 99, 104067

\bibitem[\protect\citeauthoryear{Osborne \& Jones}{Osborne \&
  Jones}{2020}]{Osborne:2019iph}
Osborne E.~L.,  Jones D.~I.,  2020, \mn@doi [Mon. Not. Roy. Astron. Soc.]
  {10.1093/mnras/staa858}, 494, 2839

\bibitem[\protect\citeauthoryear{Owen, Lindblom, Cutler, Schutz, Vecchio  \&
  Andersson}{Owen et~al.}{1998}]{Owen:1998xg}
Owen B.~J.,  Lindblom L.,  Cutler C.,  Schutz B.~F.,  Vecchio A.,   Andersson
  N.,  1998, \mn@doi [Phys. Rev. D] {10.1103/PhysRevD.58.084020}, 58, 084020

\bibitem[\protect\citeauthoryear{Palomba et~al.,}{Palomba
  et~al.}{2019}]{palomba2019direct}
Palomba C.,  et~al., 2019, \mn@doi [Phys. Rev. Lett.]
  {10.1103/PhysRevLett.123.171101}, 123, 171101

\bibitem[\protect\citeauthoryear{Piccinni et~al.,}{Piccinni
  et~al.}{2020}]{piccinni2020directed}
Piccinni O.~J.,  et~al., 2020, \mn@doi [Phys. Rev. D]
  {10.1103/PhysRevD.101.082004}, 101, 082004

\bibitem[\protect\citeauthoryear{Pierce, Riles  \& Zhao}{Pierce
  et~al.}{2018}]{PierceRilesZhao2018}
Pierce A.,  Riles K.,   Zhao Y.,  2018, \mn@doi [Phys. Rev. Lett.]
  {10.1103/PhysRevLett.121.061102}, 121, 061102

\bibitem[\protect\citeauthoryear{Pitkin, Messenger  \& Fan}{Pitkin
  et~al.}{2018}]{Pitkin:PhysRevD.98.063001}
Pitkin M.,  Messenger C.,   Fan X.,  2018, \mn@doi [Phys. Rev. D]
  {10.1103/PhysRevD.98.063001}, 98, 063001

\bibitem[\protect\citeauthoryear{Punturo et~al.}{Punturo
  et~al.}{2010}]{ET:Punturo:2010zz}
Punturo M.,  et~al., 2010, \mn@doi [Class. Quant. Grav.]
  {10.1088/0264-9381/27/19/194002}, 27, 194002

\bibitem[\protect\citeauthoryear{Reed, Deibel  \& Horowitz}{Reed
  et~al.}{2021}]{Reed:2021scb}
Reed B.~T.,  Deibel A.,   Horowitz C.~J.,  2021, \mn@doi [Astrophys. J.]
  {10.3847/1538-4357/ac1c04}, 921, 89

\bibitem[\protect\citeauthoryear{Regimbau}{Regimbau}{2011}]{Regimbau_review}
Regimbau T.,  2011, \mn@doi [Res. Astron. Astrophys.]
  {10.1088/1674-4527/11/4/001}, 11, 369

\bibitem[\protect\citeauthoryear{Reitze et~al.}{Reitze
  et~al.}{2019}]{CE:Reitze:2019iox}
Reitze D.,  et~al., 2019, Bulletin of the AAS, 51

\bibitem[\protect\citeauthoryear{Riles}{Riles}{2017}]{riles2017recent}
Riles K.,  2017, \mn@doi [Mod. Phys. Lett. A] {10.1142/S021773231730035X}, 32,
  1730035

\bibitem[\protect\citeauthoryear{Romano \& Cornish}{Romano \&
  Cornish}{2017}]{Romano:2016dpx}
Romano J.~D.,  Cornish N.~J.,  2017, \mn@doi [Living Rev. Rel.]
  {10.1007/s41114-017-0004-1}, 20, 2

\bibitem[\protect\citeauthoryear{{Rosado}}{{Rosado}}{2011}]{2011PhRvD..84h4004R}
{Rosado} P.~A.,  2011, \mn@doi [Phys. Rev. D] {10.1103/PhysRevD.84.084004},
  \href {https://ui.adsabs.harvard.edu/abs/2011PhRvD..84h4004R} {84, 084004}

\bibitem[\protect\citeauthoryear{{Rosado}}{{Rosado}}{2012}]{2012PhRvD..86j4007R}
{Rosado} P.~A.,  2012, \mn@doi [Phys. Rev. D] {10.1103/PhysRevD.86.104007},
  \href {https://ui.adsabs.harvard.edu/abs/2012PhRvD..86j4007R} {86, 104007}

\bibitem[\protect\citeauthoryear{Saleem et~al.,}{Saleem
  et~al.}{2021}]{LIGO_India:Saleem_2021}
Saleem M.,  et~al., 2021, \mn@doi [Classical and Quantum Gravity]
  {10.1088/1361-6382/ac3b99}, 39, 025004

\bibitem[\protect\citeauthoryear{Sieniawska \& Bejger}{Sieniawska \&
  Bejger}{2019}]{sieniawska2019continuous}
Sieniawska M.,  Bejger M.,  2019, \mn@doi [Universe] {10.3390/universe5110217},
  5, 217

\bibitem[\protect\citeauthoryear{Singh, Haskell, Mukherjee  \& Bulik}{Singh
  et~al.}{2020}]{Singh:2019dgy}
Singh N.,  Haskell B.,  Mukherjee D.,   Bulik T.,  2020, \mn@doi [Mon. Not.
  Roy. Astron. Soc.] {10.1093/mnras/staa442}, 493, 3866

\bibitem[\protect\citeauthoryear{Story, Gonthier  \& Harding}{Story
  et~al.}{2007}]{Story_2007}
Story S.~A.,  Gonthier P.~L.,   Harding A.~K.,  2007, \mn@doi [The
  Astrophysical Journal] {10.1086/521016}, 671, 713

\bibitem[\protect\citeauthoryear{Sun \& Melatos}{Sun \&
  Melatos}{2019}]{Sun:2018hmm}
Sun L.,  Melatos A.,  2019, \mn@doi [Phys. Rev. D]
  {10.1103/PhysRevD.99.123003}, 99, 123003

\bibitem[\protect\citeauthoryear{Sun, Brito  \& Isi}{Sun
  et~al.}{2020}]{Sun:2019mqb}
Sun L.,  Brito R.,   Isi M.,  2020, \mn@doi [Phys. Rev. D]
  {10.1103/PhysRevD.101.063020}, 101, 063020

\bibitem[\protect\citeauthoryear{Suresh, Ain  \& Mitra}{Suresh
  et~al.}{2021}]{pystochSpH}
Suresh J.,  Ain A.,   Mitra S.,  2021, \mn@doi [Phys. Rev. D]
  {10.1103/PhysRevD.103.083024}, 103, 083024

\bibitem[\protect\citeauthoryear{Talukder, Mitra  \& Bose}{Talukder
  et~al.}{2011}]{lambda_statistic}
Talukder D.,  Mitra S.,   Bose S.,  2011, \mn@doi [Phys. Rev. D]
  {10.1103/PhysRevD.83.063002}, 83, 063002

\bibitem[\protect\citeauthoryear{Talukder, Thrane, Bose  \& Regimbau}{Talukder
  et~al.}{2014}]{Talukder:2014eba}
Talukder D.,  Thrane E.,  Bose S.,   Regimbau T.,  2014, \mn@doi [Phys. Rev. D]
  {10.1103/PhysRevD.89.123008}, 89, 123008

\bibitem[\protect\citeauthoryear{Tenorio, Keitel  \& Sintes}{Tenorio
  et~al.}{2021}]{Tenorio:2021wmz}
Tenorio R.,  Keitel D.,   Sintes A.~M.,  2021, \mn@doi [Universe]
  {10.3390/universe7120474}, 7, 474

\bibitem[\protect\citeauthoryear{Thrane \& Romano}{Thrane \&
  Romano}{2013}]{Thrane_Romano_sensitivities_PhysRevD.88.124032}
Thrane E.,  Romano J.~D.,  2013, \mn@doi [Phys. Rev. D]
  {10.1103/PhysRevD.88.124032}, 88, 124032

\bibitem[\protect\citeauthoryear{Thrane, Ballmer, Romano, Mitra, Talukder, Bose
   \& Mandic}{Thrane et~al.}{2009}]{Thrane:2009}
Thrane E.,  Ballmer S.,  Romano J.~D.,  Mitra S.,  Talukder D.,  Bose S.,
  Mandic V.,  2009, \mn@doi [Phys. Rev. D] {10.1103/PhysRevD.80.122002}, 80,
  122002

\bibitem[\protect\citeauthoryear{Turner}{Turner}{1997}]{Turner}
Turner M.~S.,  1997, \mn@doi [Phys. Rev. D] {10.1103/PhysRevD.55.R435}, 55,
  R435

\bibitem[\protect\citeauthoryear{Ushomirsky, Cutler  \& Bildsten}{Ushomirsky
  et~al.}{2000}]{Ushomirsky:2000ax}
Ushomirsky G.,  Cutler C.,   Bildsten L.,  2000, \mn@doi [Mon. Not. Roy.
  Astron. Soc.] {10.1046/j.1365-8711.2000.03938.x}, 319, 902

\bibitem[\protect\citeauthoryear{Virtanen et~al.,}{Virtanen
  et~al.}{2020}]{2020SciPy-NMeth}
Virtanen P.,  et~al., 2020, \mn@doi [Nature Methods]
  {10.1038/s41592-019-0686-2}, \href {https://rdcu.be/b08Wh} {17, 261}

\bibitem[\protect\citeauthoryear{Watanabe \& Komatsu}{Watanabe \&
  Komatsu}{2006}]{Watanabe}
Watanabe Y.,  Komatsu E.,  2006, \mn@doi [Phys. Rev. D]
  {10.1103/PhysRevD.73.123515}, 73, 123515

\bibitem[\protect\citeauthoryear{{Wu}, {Mandic}  \& {Regimbau}}{{Wu}
  et~al.}{2013}]{2013PhRvD..87d2002W}
{Wu} C.-J.,  {Mandic} V.,   {Regimbau} T.,  2013, \mn@doi [Phys. Rev. D]
  {10.1103/PhysRevD.87.042002}, \href
  {https://ui.adsabs.harvard.edu/abs/2013PhRvD..87d2002W} {87, 042002}

\bibitem[\protect\citeauthoryear{{Zhu}, {Howell}, {Regimbau}, {Blair}  \&
  {Zhu}}{{Zhu} et~al.}{2011}]{2011ApJ73986Z}
{Zhu} X.-J.,  {Howell} E.,  {Regimbau} T.,  {Blair} D.,   {Zhu} Z.-H.,  2011,
  \mn@doi [Astrophys. J.] {10.1088/0004-637X/739/2/86}, \href
  {https://ui.adsabs.harvard.edu/abs/2011ApJ...739...86Z} {739, 86}

\bibitem[\protect\citeauthoryear{{Zhu}, {Howell}, {Blair}  \& {Zhu}}{{Zhu}
  et~al.}{2013}]{2013MNRAS.431..882Z}
{Zhu} X.-J.,  {Howell} E.~J.,  {Blair} D.~G.,   {Zhu} Z.-H.,  2013, \mn@doi
  [Mon. Not. Roy. Astron. Soc.] {10.1093/mnras/stt207}, \href
  {https://ui.adsabs.harvard.edu/abs/2013MNRAS.431..882Z} {431, 882}

\bibitem[\protect\citeauthoryear{Zonca, Singer, Lenz, Reinecke, Rosset, Hivon
  \& Gorski}{Zonca et~al.}{2019}]{Healpy:Zonca2019}
Zonca A.,  Singer L.~P.,  Lenz D.,  Reinecke M.,  Rosset C.,  Hivon E.,
  Gorski K.~M.,  2019, \mn@doi [Journal of Open Source Software]
  {10.21105/joss.01298}, 4, 1298

\makeatother
\end{thebibliography}

\appendix

\section{Derivation of the spectral shape}
\label{app:H(f)}
Here, we present a heuristic procedure to get the expression of the spectral shape $H(f)$ in equation \eqref{eq:spectral_shape}. It is known that the GW power emitted by a pulsar at a given frequency can be expressed as \citep{maggiore2008gravitational}:
\begin{equation}
\label{eq:pulsar_power}
    P(f) = \frac{32 \, \upi^6 \, G}{5 \,c^5}\varepsilon^2 \, I^2_{zz} \,f^6\,,
\end{equation}
and is linked to the GW energy density at a distance $d$ from the source through the relation
\begin{equation}
    \rho_{\mathrm{gw}}(f)=\frac{P(f)}{4\upi \,c\, d^2}.
\end{equation}
Now, let us consider a population of NSs, each one emitting GW at a frequency $f_j$ and from the direction $\hat{n}_j$, where ${j = 1, 2, ..., N_{0}}$ and $N_0$ is the total number of pulsars. Then, the corresponding GW energy density ratio can be expressed as
\begin{equation}
    \label{eq:omega_pulsar}
    \begin{split}
    \Omega_{\mathrm{gw}}(f, \hat{n}) &= \frac{f}{\rho_c}\sum_{j=1}^{N_0} \frac{P_j}{4\upi c d_j^2} \,\delta(f-f_j) \, \delta^2 (\hat{n}, \hat{n}_j) \\
    &= \frac{f \left\langle P \right\rangle_{\mathrm{NS}}}{4 \upi \rho_{c} \, c} \left\langle\frac{1}{d^2}\right\rangle_{\mathrm{NS}} N_0 \, \Psi(f, \hat{n}) \,,
    \end{split}
\end{equation}
where $\Psi(f, \hat{n})$ is the frequency-angular distribution of the NS population. We take the ensemble average over pulsars parameters in the second line of above equation.
Finally, plugging equation \eqref{eq:pulsar_power} in equation \eqref{eq:omega_pulsar} and using the definition of spectral shape (equation \eqref{eq:omega_vs_shape}), we get
\begin{equation}
   \begin{split}
    H(f) &= \frac{3 H_0^2}{2\upi^2 f^3}\int d^2\hat{n} \, \Omega_{\mathrm{gw}}(f, \hat{n})\\
    &= \frac{32 \upi^4 \, G^2 \left\langle \varepsilon ^2 \right\rangle_{\mathrm{NS}} \left\langle I_{zz}^2 \right\rangle_{\mathrm{NS}}}{5 c^8 \left\langle d^2 \right\rangle_{\mathrm{NS}}} f^4 \, \Phi(f) \, N_0 \,,
    \end{split}
\end{equation}
this is identical to equation \eqref{eq:spectral_shape} where $\Phi(f) \equiv \int d^2\Omega_{\hat{n}} \,\Psi(f, \hat{n})$.

\section{The ellipticity estimator uncertainty: general case}
\label{app:sigma_eps}
Here, we derive the expression of the variance $\sigma^2_{\hat{\varepsilon}}(f_{\mathrm{k}})$ of the estimator of the average ellipticity $\hat{\varepsilon}_{\mathrm{av}}(f_{\mathrm{k}})$ in the general case, where $\hat{\varepsilon}_{\mathrm{av}}(f_{\mathrm{k}})>0$.
The procedure is not complex conceptually: it is necessary to evaluate the first and second-order expectation values starting from the likelihood in equation \eqref{eq:lklhood_epsilon} and combine them to get $\sigma^2_{\hat{\varepsilon}}(f_{\mathrm{k}})$, but some algebra is required. 

From the expressions (where $D_{\nu}(z)$ is a parabolic cylinder function and we omit the frequency label in the last expression)
\begin{equation}
    \begin{split}
    \left\langle \varepsilon_{\mathrm{av}} (f_{\mathrm{k}})\right\rangle &\equiv \int_{0}^{\infty} p_{\varepsilon} \left(\hat{\varepsilon}_{\mathrm{av}}(f_{\mathrm{k}})|\varepsilon_{\mathrm{av}}(f_{\mathrm{k}})\right) \varepsilon_{\mathrm{av}} \, d\varepsilon_{\mathrm{av}} \\
    &= \sqrt{\frac{\sigma_{\hat{\Omega}}}{2\xi}} D_{-3/2}\left(-\frac{\hat{\varepsilon}_{\mathrm{av}}^2 \, \xi}{\sigma_{\hat{\Omega}}}\right)\exp\left[{-\frac{\hat{\varepsilon}_{\mathrm{av}}^4 \, \xi^2}{4\sigma_{\hat{\Omega}}^2}}\right]
    \end{split}
\end{equation}
and
\begin{equation}
\begin{split}
    \left\langle \varepsilon_{\mathrm{av}}^2 (f_{\mathrm{k}})\right\rangle &\equiv \int_{0}^{\infty} p_{\varepsilon} \left(\hat{\varepsilon}_{\mathrm{av}}(f_{\mathrm{k}})|\varepsilon(_{\mathrm{av}}f_{\mathrm{k}})\right) \varepsilon_{\mathrm{av}}^2 \, d\varepsilon_{\mathrm{av}} \\
    &= \sqrt{\frac{2}{\upi}}\frac{\sigma_{\hat{\Omega}}}{\xi}D_{-2}\left(-\frac{\hat{\varepsilon}_{\mathrm{av}}^2 \, \xi}{\sigma_{\hat{\Omega}}}\right) \exp\left[-\frac{\hat{\varepsilon}_{\mathrm{av}}^4 \, \xi^2}{4\sigma_{\hat{\Omega}}^2}\right]\,,
\end{split}
\end{equation}
it is straightforward to show
\begin{equation}
    \label{eq:sigma_epsilon_full}
    \begin{split}
    \sigma^2_{\hat{\varepsilon}} (f_{\mathrm{k}}) =& \left\langle \varepsilon_{\mathrm{av}}^2 (f_{\mathrm{k}}) \right\rangle - \left\langle \varepsilon_{\mathrm{av}} (f_{\mathrm{k}}) \right\rangle^2 \\
    =&\sqrt{\frac{2}{\upi}} \frac{\sigma_{\hat{\Omega}}}{\xi} D_{-2}\left(-\frac{\hat{\varepsilon}_{\mathrm{av}}^2\,\xi}{\sigma_{\hat{\Omega}}}\right) \exp\left[-\frac{\hat{\varepsilon}_{\mathrm{av}}^4\,\xi^2}{4\sigma_{\hat{\Omega}}^2}\right] - \\
    &- \frac{1}{2} \frac{\sigma_{\hat{\Omega}}}{\xi} D^2_{-3/2}\left(-\frac{\hat{\varepsilon}_{\mathrm{av}}^2\,\xi}{\sigma_{\hat{\Omega}}}\right)\exp\left[-\frac{\hat{\varepsilon}_{\mathrm{av}}^4\,\xi^2}{2\sigma_{\hat{\Omega}}^2}\right]\,.
    \end{split}
\end{equation}
The limit $\varepsilon << 1$ in equation \eqref{eq:sigma_epsilon} is recovered by observing and using 
\begin{equation}
    D_{\nu}(z)|_{z<<1} \approx \frac{2^{-(\nu+2)/2} \, \Gamma\left(-\frac{\nu}{2}\right)}{\Gamma(-\nu)}\,,
\end{equation}
where $\Gamma(z)$ is Euler's Gamma function.

\section{Search for SGWB from NS hotpots}
\label{app:BBR}
\begin{figure}
    \centering
    \includegraphics[width = \columnwidth]{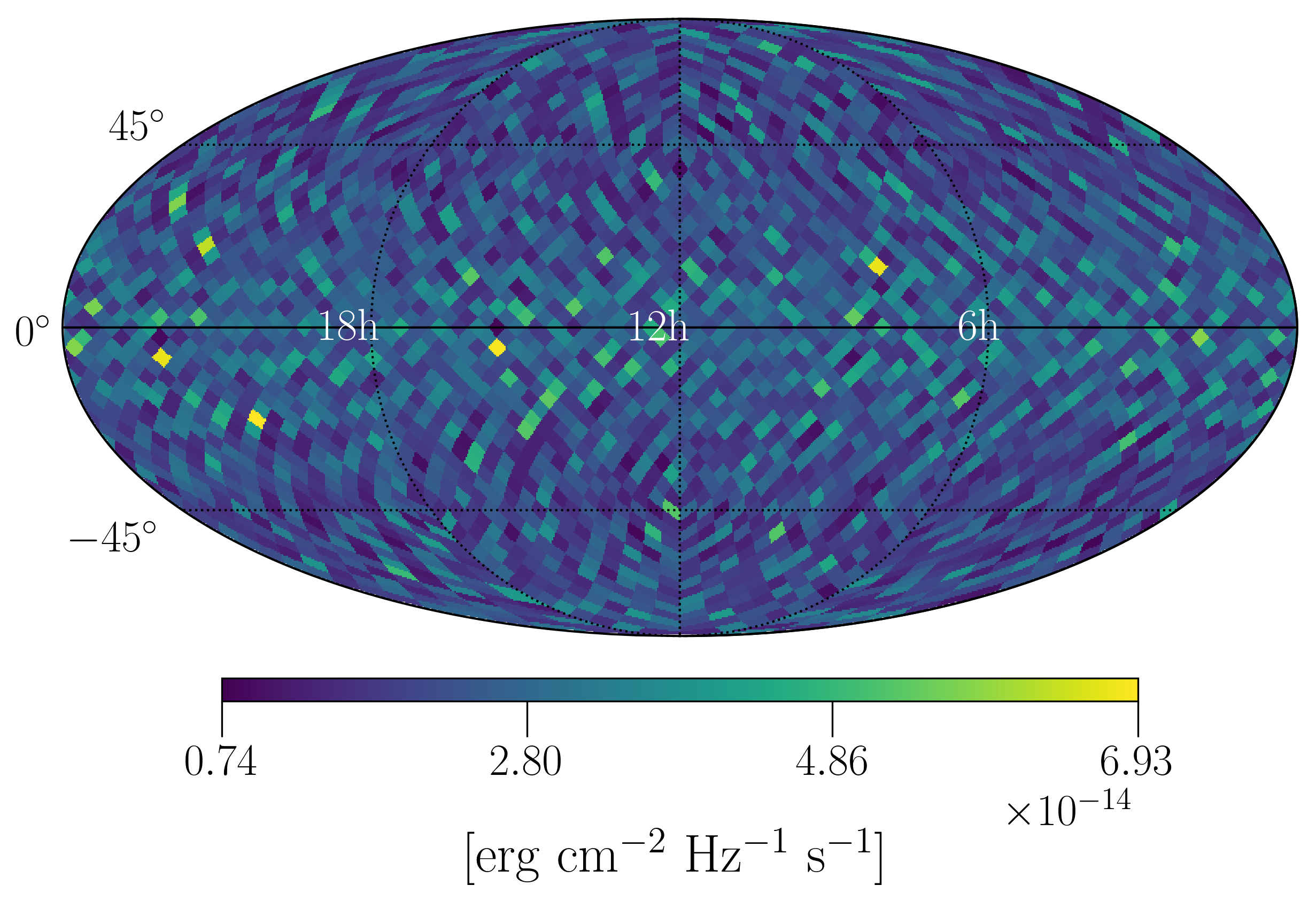}
    \caption{Upper limit sky maps on GW energy flux from the broadband-radiometer analysis for the model $\bar{H}(f)$ in equation \eqref{eq:H_bar}. Here the NSs frequency distribution $\Phi(f)$ is the one built from the ATNF catalogue as described in section \ref{Sec:Model}. The sky map is represented as a color bar plot on a Mollweide projection of the sky in ecliptic coordinates with $N_{\mathrm{side}} = 16$.}
    \label{fig:BBR_flux_upper_limits}
\end{figure}
Here, we present how we have derived the limits on $\Omega_{\mathrm{gw}}^{\mathrm{patch}} (f)$ of a patch in the sky to use as input in section \ref{sec:hotspot_results} to get the constraints on the average ellipticity of the NS hotspots.

\subsection{The directional radiometer search}
The directional radiometer search drops the assumption of the SGWB being isotropic \citep{PhysRevD.98.063509,2014PhRvD..89h4076M,PhysRevLett.120.231101, 2018PhRvD..98f3501J,2012PhRvD..86j4007R, 2013PhRvD..87d2002W,2013PhRvD..87f3004L,PhysRevD.96.103019}.
This means that the background cannot be simply characterised by considering $\Omega_{\mathrm{gw}}(f)$ in equation \eqref{eq:omega_f}, but rather the frequency-angular dependent density parameter $\Omega_{\mathrm{gw}}(f, \hat{n})$ (measured in $\mathrm{sr^{-1}}$):
 \begin{equation}
     \label{eq:omega_f_n}
     \Omega_{\mathrm{gw}}(f, \hat{n}) = \frac{f}{\rho_{c}} \, \frac{d^3\rho_{\mathrm{gw}}(f, \hat{n})}{df d^2\hat{n}}= \frac{2\upi^2}{3H_0^2}\, f^3 \, \mathcal{P}(f, \hat{n}) \,,
 \end{equation}
with $\mathcal{P}(f, \hat{n})$ being the GW strain power.

To measure the anisotropies, the radiometer search introduces a maximum-likelihood (ML) estimator \citep{MitraML:PhysRevD.77.042002,Thrane:2009}, as statistic, at each frequency and each direction \citep{ASAF:LIGOScientific:2021oez} $\hat{\mathcal{P}}(f, \hat{n})$ with cross-correlation matrix $\sigma_{\hat{n}, \hat{n}'}(f)$:
\begin{align}
    \label{eq:clean_map}
    \hat{\mathcal{P}}(f, \hat{n}) &= \sum_{\hat{n}'}\left[\Gamma_{\hat{n}\hat{n}'}(f)\right]^{-1} X_{\hat{n}'}(f), \\
    \sigma_{\hat{n}, \hat{n}'}(f) &= \left[\Gamma_{\hat{n}\hat{n}'}(f)\right]^{-1/2},
\end{align}
where $X_{\hat{n}'}(f)$ is called ``dirty map'' and $\Gamma_{\hat{n}\hat{n}'}$ is the Fisher information matrix in the small-signal limit. The summation over $\hat{n}'$ implies integration over the solid angle. The dirty map represents the sky seen through the response of a set of independent baselines $IJ$, defined as
\begin{equation}
    X_{\hat{n}}(f) = \tau \Delta f \, \Re \sum_{IJ, t} \, \frac{\left[\gamma_{IJ}(t; f)\right]^{*}_{\hat{n}} \,\hat{C}_{IJ}(t;f)}{P_{I}(t;f) \, P_{J}(t;f)},
\end{equation}
where $\hat{C}_{IJ}(t;f) \equiv (2/\tau)\,\tilde{s}^{*}_{I} (t; f) \,\tilde{s}_{J} (t;f)$ is the cross-correlation spectral density, while $\gamma_{IJ}(t; f, \hat{n})$ is the directional overlap reduction function, which is proportional to the isotropic one in equation \eqref{eq:CC_iso} when integrated over the sky. The Fisher information matrix encodes the uncertainty in the measurement of the dirty map, and is defined as
\begin{equation}
    \Gamma_{\hat{n}, \hat{n}'} (f) = \tau \Delta f \, \Re \sum_{IJ, t} \, \frac{\left[\gamma_{IJ}(t; f)\right]^{*}_{\hat{n}} \, \left[\gamma_{IJ}(t; f)\right]_{\hat{n}'}}{P_{I}(t;f) \, P_{J}(t;f)}.
\end{equation}

The ML estimator $\hat{\mathcal{P}}(f, \hat{n})$ in equation \eqref{eq:clean_map}, involves the inversion of $\Gamma_{\hat{n}, \hat{n}'}(f)$, which can be singular in general and must be regularised. However, for point-like sources considered here, we can work by employing the pixel basis
\begin{equation}
    \label{eq:pixel_basis}
    \mathcal{P}(f, \hat{n}) \equiv \mathcal{P}(f, \hat{n}') \, \delta^2(\hat{n}, \hat{n}'),
\end{equation}
and ignore the correlation among neighbourhood directions in the sky \citep{O3_Aniso_PhysRevD.104.022005, ASAF:LIGOScientific:2021oez}, and the Fisher information matrix is no longer singular and becomes diagonal. With this caveat, the estimator can be used to set upper limits on $\Omega_{\mathrm{gw}}(f, \hat{n})$ and related quantities.

\subsection{SGWB from NS hotspots in the sky}
To specialise the framework to our search for a SGWB from NS hotspots, we make the following, standard, ansatz about the factorisability of $\hat{\mathcal{P}}(f, \hat{n})$ in a frequency and direction dependent only terms:
\begin{equation}
    \label{eq:P_n_factorisation}
    \mathcal{P}(f, \hat{n}) = \bar{H}(f)\,\mathcal{P}(\hat{n}) \,,
\end{equation}
where $\bar{H}(f)$ is defined in such a way that $\bar{H}(f_{\mathrm{ref}})=1$, $\mathcal{P}(\hat{n})$ is the angular distribution of gravitational-wave power to be estimated by the search. For the signal model presented in section \ref{Sec:Methods}, $\bar{H}(f)$ turns out to be
\begin{equation}
    \label{eq:H_bar}
    \bar{H}(f) = \left(\frac{f}{f_{\mathrm{ref}}}\right)^{4}\frac{\Phi(f)}{\Phi(f_{\mathrm{ref}})} \,.
\end{equation}

Using the above formalism, it is possible to integrate $\hat{\mathcal{P}}(f, \hat{n})$ (or, equivalently $\hat{\Omega}_{\mathrm{gw}}(f,\hat{n})$) over the frequencies, to get a set of broadband estimators $\hat{\mathcal{P}}_{\mathrm{ref}}(\hat{n})$, or over the sky directions of a sky patch, to get a set of narrowband estimators $\hat{\mathcal{P}}^{\mathrm{patch}}(f_{\mathrm{k}})$, or even both frequencies and direction, getting a broadband estimator of the SGWB of a sky patch $\hat{\mathcal{P}}_{\mathrm{ref}}^{\mathrm{patch}}$. The master formulas for the integrated estimator over a set of frequencies and directions, and the relative uncertainty, is given by
\begin{align}
    \hat{\mathcal{P}}_{\mathrm{ref}}^{\mathrm{patch}} &=
    \frac{\sum_{k, j}\hat{\mathcal{P}}(f_{\mathrm{k}}, \hat{n}_\mathrm{j})\,\sigma^{-2}(f_{\mathrm{k}}, \hat{n}_\mathrm{j}) \, \bar{H}(f)}{\sum_{k, j}\sigma^{-2}(f_{\mathrm{k}}, \hat{n}_\mathrm{j}) \, \bar{H}(f)^2},\\
    \sigma_{\mathrm{ref}}^{\mathrm{patch}} &= \left(\sum_{k, j}\sigma^{-2}(f_{\mathrm{k}}, \hat{n}_\mathrm{j}) \,  \bar{H}(f)^2\right)^{-1/2}.
\end{align}

Within this framework, we derive the following quantities: a broadband estimator for each sky direction $\hat{\mathcal{P}}_{\mathrm{ref}}^{\mathrm{sky}}(\hat{n})$, a set of narrowband estimators for each patch $\hat{\mathcal{P}}^{\mathrm{patch}}(f_{\mathrm{k}})$, and a broadband estimator for each patch $\hat{\mathcal{P}}_{\mathrm{ref}}^{\mathrm{patch}}$. The broadband estimators are evaluated to get information about the SGWB from a NS population, when one allows for an unknown spatial distribution of the population, compared to the isotropic case. These estimators are translated into estimators of the GW energy flux, given the astrophysical nature of the source, 
\begin{equation}
    \label{eq:oflux_hat_directional}
    \hat{\mathcal{F}}_{\mathrm{ref}}(\hat{n}) = \frac{c^3 \upi}{4 G} \, f_{\mathrm{ref}}^2 \, \hat{\mathcal{P}}_{\mathrm{ref}}(\hat{n}),
\end{equation}
at a reference frequency $f_{\mathrm{ref}}$, from which the relative UL are calculated and illustrated in figure \ref{fig:BBR_flux_upper_limits}. The narrowband estimators for every patch, instead, are converted to narrowband estimators of the density parameter $\hat{\Omega}_{\mathrm{gw}}^{\mathrm{patch}}(f_{\mathrm{k}})$ by means of equation \eqref{eq:omega_f_n}, and are used as input for the evaluation of constraints on the average ellipticity of the different NS populations in section \ref{sec:hotspot_results}. 

\bsp	
\label{lastpage}
\end{document}